\documentclass[12pt,letterpaper,usenames,dvipsnames]{article}
\usepackage{jheppub}
\usepackage{mathrsfs} 
\usepackage{enumitem} 
\usepackage{amsthm} 
\usepackage{dsfont} 
\usepackage{physics} 

\theoremstyle{definition}

\newtheorem{example}{Example}[section]
\newtheorem{theorem}{Theorem}[section]

\usepackage{outlines} 
\newcommand{\bo}{\begin{outline}} 
\newcommand{\eo}{\end{outline}}

\usepackage{tikz}
\usepackage{tikz-cd} 
\usetikzlibrary{cd, arrows, calc, shapes, shapes.arrows, decorations.markings, decorations.pathmorphing}

\usepackage{multirow} 
\usepackage{colortbl} 
\usepackage{arydshln} 
\usepackage{graphicx} 
\usepackage{float} 
\usepackage{blkarray} 
\usepackage{tabularx} 
\usepackage{booktabs}

\definecolor{airforceblue}{rgb}{0.36, 0.54, 0.66}

\definecolor{lgray}{RGB}{240, 240, 240}

\definecolor{mDarkTeal}{HTML}{23373b}
\definecolor{tred}{HTML}{E00122}

\definecolor{lred}{HTML}{ff5b59}

\definecolor{teal}{HTML}{0097AD}

\definecolor{seagreen}{HTML}{009382}

\definecolor{water}{HTML}{239CFF}

\definecolor{lblue}{HTML}{00C7E9}

\definecolor{dblue}{HTML}{014FFF}

\definecolor{org}{HTML}{FD6E00}

\definecolor{lorg}{HTML}{FFB887}
\definecolor{gfruit}{HTML}{ff5b59}

\definecolor{plum}{HTML}{5B83FF}
\definecolor{airforceblue}{rgb}{0.36, 0.54, 0.66}

\definecolor{rem}{HTML}{0097AD}

\def\deq{\doteq}
\def\til{\widetilde}
\def\wh{\widehat}

\def\^{\wedge}
\def\bar{\overline}

\makeatletter 
 \newcommand{\bw}{\@ifnextchar^\@bw{\@bw^{\,}}}
 \def\@bw^#1{\mathop{\bigwedge\nolimits^{\!#1}}}
\makeatother
\def\<{\langle}
\def\>{\rangle}
\def\del{{\partial}}
\def\delb{{\bar\del}}

\def\nn{\nonumber}
\def\ph{\phantom}
\newcommand{\bpm}{\begin{pmatrix}}
\newcommand{\epm}{\end{pmatrix}}
\newcommand{\bsm}{\begin{smallmatrix}} 
\newcommand{\esm}{\end{smallmatrix}}
\newcommand{\bspm}{\left(\begin{smallmatrix}}
\newcommand{\espm}{\end{smallmatrix}\right)}

\def\Im{{\rm Im}}

\def\Aut{{\text{Aut}}}
\def\diag{{\text{diag}}}
\def\diff{{\mathrm d}}

\def\GL{{\rm GL}}

\def\su{\mathfrak{su}}
\def\Sp{{\rm Sp}}

\def\U{{\rm U}}

\def\sfA{{\mathsf A}}

\def\sfB{{\mathsf B}}
\def\cC{{\mathcal C}}
\def\C{{\mathbb C}}

\def\cF{{\mathcal F}}

\def\cH{{\mathcal H}}

\def\cL{{\mathcal L}}

\def\N{{\mathbb N}} 
\def\cN{{\mathcal N}}
 
\def\Q{{\mathbb Q}} 

\def\R{{\mathbb R}}

\def\hx{{\hat x}}
\def\tx{{\til x}}
\def\htx{{\wh\tx}}
\def\cX{{\mathcal X}}

\def\Z{{\mathbb Z}}
\def\fz{{\mathfrak z}}
\def\zb{{\bar z}}
\def\tz{{\til z}}

\def\a{{\alpha}}
\def\ta{{\til\a}}

\def\b{{\beta}}
\def\tb{{\til\b}}

\def\g{{\gamma}}

\def\d{{\delta}}

\def\D{{\Delta}}
\def\e{{\epsilon}}

\def\z{{\zeta}}
\def\th{{\theta}}
\def\thb{{\bar\th}}
\def\tth{{\til\th}}

\def\k{{\kappa}}
\def\l{{\lambda}}
\def\tl{{\til\l}}
\def\L{{\Lambda}}
\def\m{{\mu}}
\def\tm{{\til\m}}

\def\r{{\rho}}
\def\s{{\sigma}}

\def\t{{\tau}}
\def\bt{{\bar\t}}

\def\w{{\omega}}
\def\tw{{\til\w}}
\def\Om{{\Omega}}

\DeclareFontFamily{U}{wncy}{}
\DeclareFontShape{U}{wncy}{m}{n}{<->wncyr10}{}
\DeclareSymbolFont{mcy}{U}{wncy}{m}{n}
\DeclareMathSymbol{\Sh}{\mathord}{mcy}{"58} 

\title{Completely (iso-)split scale-invariant Coulomb branch geometries are isotrivial}
\author[a]{Philip C. Argyres,}
\author[b]{Robert Moscrop,}
\author[a]{Souradeep Thakur,}
\author[c]{Mitch Weaver}
\affiliation[a]{Physics Department, University of Cincinnati, PO Box 210011, Cincinnati OH 45221 USA}
\affiliation[b]{Center  of  Mathematical  Sciences  and  Applications,  Harvard  University,  MA  02138,  USA}
\affiliation[c]{Department of Physics, Korea Advanced Institute of Science and Technology 291 Daehak-ro, Yuseong-gu, Daejeon 34141, Republic of Korea}
\emailAdd{philip.argyres@gmail.com}
\emailAdd{robert@cmsa.fas.harvard.edu}
\emailAdd{thakursp@mail.uc.edu}
\emailAdd{mtw7497@gmail.com}

\abstract{We show that scale-invariant special K\"ahler geometries whose generic $r$ $\dim_\C$ abelian variety fiber is isomorphic (completely split) or isogenous (completely iso-split) as a complex torus to the product of $r$ one-dimensional complex tori have constant $\t^{ij}$ modulus on the Coulomb branch, i.e. are isotrivial.
These simple results are useful in organizing the classification of scale-invariant special K\"ahler geometries, which, in turn, is relevant to the classification of possible 4-dimensional $\cN=2$ supersymmetric superconformal field theories.
}

\begin{document}

\maketitle

\section{Introduction}\label{section 1}

Superconformal field theories (SCFTs) with 8 or more Poincar\'e supercharges are very tightly constrained, and seem to form a denumerable family of field theories.  
In four space-time dimensions the Coulomb branch (CB) of the moduli space of $\cN=2$ SCFTs are special K\"ahler (SK) varieties with a complex scale symmetry which acts freely except at the superconformal vacuum.
These SK varieties are stratified algebraic varieties whose strata are smooth SK manifolds \cite{Freed:1997dp, Martone:2020nsy, Argyres:2020wmq},
and the resulting CB geometries are rigid by virtue of their discrete electric-magnetic (EM) monodromy groups \cite{Cecotti:2023ksl}.
This motivates the attempt to classify scale-invariant CB geometries.
Physically, such a classification puts a priori (``bottom up'') constraints on the space of SCFTs.
More practically, comparison between even partial results in this geometric classification with field- and string-theoretic constructions of SCFTs will inform our understanding of what stratification structures are allowed in physical SCFTs.
This will further refine our understanding of what singular behaviors can occur along the intersections of co-dimension 1 and higher singular strata within rank $r>1$ CBs.%
\footnote{The rank of a CB is its complex dimension.}

The low-energy theory around a generic point on a rank-$r$ CB is captured by a fibration of rank-$r$ abelian varieties \cite{Donagi94, Seiberg:1994aj, Donagi:1995cf}.
We will call CB geometries whose generic abelian variety fiber is a direct product of lower-dimensional tori, ``split" geometries.
Note that with our definition, a split CB geometry need not factorize as a direct product of lower-rank CB geometries even though the total space of the complex torus fibration factorizes as a fiber product.
A subset of split CB geometries are ``completely split'' geometries: ones in which the generic fiber completely factorizes as a direct product of $1$-dimensional complex tori.
Note that at rank $2$, all split SK geometries are completely split.

In this note we show in an elementary way that completely split SK geometries are \emph{isotrivial} if the geometry is scale invariant.
Isotrivial SK geometries are ones in which the complex modulus $\t^{ij}$ of the SK structure --- identified with the matrix of complex EM couplings in the low energy theory on the CB --- is constant.
This observation about the local structure of CBs is useful since there are strong classification results concerning global isotrivial geometries \cite{Cecotti:2021ouq, ACDMMTW24}. 
Our results on completely (iso)-split geometries apply regardless of the value of the characteristic dimension invariant, $\varkappa$, for the CB geometry defined in \cite{Cecotti:2021ouq}.
In particular, our derivation of isotriviality applies for theories that are completely (iso)-split and have $\varkappa \in \{ 1,2\}$, where the arguments of \cite{Cecotti:2021ouq} do not imply isotriviality.

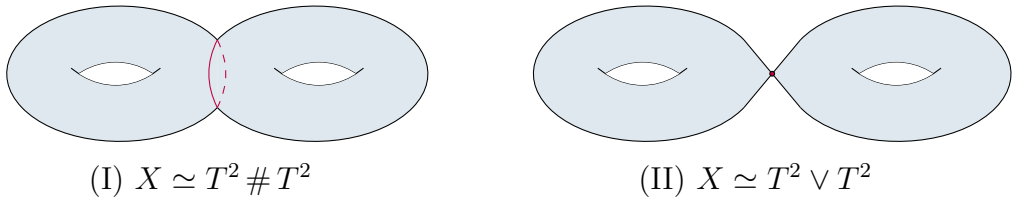
\begin{figure}
    \centering\vspace{-0cm}
    \begin{tikzpicture}[scale=0.75]
        \draw[fill=airforceblue!20] (0,0) arc[start angle=30, end angle=330, x radius=2cm, y radius=1.2cm];
        \begin{scope}[xscale=-1]
            \draw[fill=airforceblue!20] (0,0) arc[start angle=30, end angle=330, x radius=2cm, y radius=1.2cm];
        \end{scope}
        \begin{scope}
            \clip (-1.8,-2.2) ellipse (1.5 and 2);
            \draw (-1.8,1.2) ellipse (1.5 and 2);
            \clip (-1.8,1.2) ellipse (1.5 and 2);
            \draw (-1.8,-2.4) ellipse (1.5 and 2);
            \fill[white] (-1.8,-2.4) ellipse (1.5 and 2);
        \end{scope}
        \begin{scope}[xscale=-1]
            \clip (-1.8,-2.2) ellipse (1.5 and 2);
            \draw (-1.8,1.2) ellipse (1.5 and 2);
            \clip (-1.8,1.2) ellipse (1.5 and 2);
            \draw (-1.8,-2.4) ellipse (1.5 and 2);
            \fill[white] (-1.8,-2.4) ellipse (1.5 and 2);
        \end{scope}
        \draw[color=purple, dashed] (0,0) .. controls (0.2, -0.4) and (0.2,-0.8) .. (0,-1.2);
        \draw[color=purple] (0,0) .. controls (-0.2, -0.4) and (-0.2,-0.8) .. (0,-1.2);
        \node[below] at (-0.3,-2){(I) $X\simeq T^2\, \#\, T^2$};
    \end{tikzpicture}
    \quad\quad
        \begin{tikzpicture}[scale=0.75]
        \draw[color=airforceblue!20] (0,0) -- (0,-1.2);
        \draw[color=airforceblue!20] (1,0) -- (1,-1.2);
        \draw[fill=airforceblue!20] (0,0) arc[start angle=30, end angle=330, x radius=2cm, y radius=1.2cm];
        \begin{scope}[xscale=-1]
            \draw[fill=airforceblue!20] (-1,0) arc[start angle=30, end angle=330, x radius=2cm, y radius=1.2cm];
        \end{scope}
        \begin{scope}
            \clip (-1.8,-2.2) ellipse (1.5 and 2);
            \draw (-1.8,1.2) ellipse (1.5 and 2);
            \clip (-1.8,1.2) ellipse (1.5 and 2);
            \draw (-1.8,-2.4) ellipse (1.5 and 2);
            \fill[white] (-1.8,-2.4) ellipse (1.5 and 2);
        \end{scope}
        \begin{scope}[xscale=-1]
            \clip (-2.7,-2.2) ellipse (1.5 and 2);
            \draw (-2.7,1.2) ellipse (1.5 and 2);
            \clip (-2.7,1.2) ellipse (1.5 and 2);
            \draw (-2.7,-2.4) ellipse (1.5 and 2);
            \fill[white] (-2.7,-2.4) ellipse (1.5 and 2);
        \end{scope}
        \draw[fill=airforceblue!20] (0,0) -- (0.5,-0.6)  -- (0,-1.2);
        \draw[fill=airforceblue!20] (1,0) -- (0.5,-0.6) -- (1,-1.2);
        \draw[fill=purple] (0.5,-0.6) circle (1.2pt);
         \node[below] at (0.2,-2){(II) $X\simeq T^2 \vee T^2$};
    \end{tikzpicture}\vspace{-1cm}
    \caption{{\it Left:} A generic non-degenerate Riemann surface of genus 2.
    Their Jacobian varieties are always irreducible principally polarized abelian varieties, but in special cases they may split as a complex torus.
    {\it Right:} A degenerate genus 2 Riemann surface. 
    These can be understood as a bouquet of two genus 1 tori. 
   The Jacobian varieties of the latter are always smooth, may split as a polarized abelian variety, and will always give rise to a split complex torus.
   }
    \label{fig:2tori}
\end{figure}

We now outline a derivation of our results.
SK geometry is described locally in terms of the total space of a holomorphic fibration of rank-$r$ abelian varieties over an open, smooth, contractible subset, $\cC$, of the rank-$r$ CB, together with a holomorphic symplectic form, $L$, on the total space with respect to which the fibers are lagrangian.
If completely split, then the fibration factorizes as the fiber product of $r$ rank-$1$ torus fibrations over $\cC$.
In particular, there is a $1$-homology and $(1,0)$-cohomology basis of the rank-$r$ abelian variety fiber in which its $r\times 2r$ period matrix, $\Pi(u)$, over any point $u\in \cC$ takes the $r\times r$ block matrix form $\Pi(u) = (\mathds 1_{r}\ Z(u))$ where the modulus, $Z^{ij}$, is diagonal,
\begin{align}\label{c split1}
    Z^{ij}(u) &= 
    \bpm Z^1(u) & & \\ & \ddots & \\ && Z^r(u) \epm ,&\quad \Im \, Z^i(u) &> 0 , 
\end{align}
and each rank-$1$ modulus $Z^i(u)$ is a priori a non-constant function of $u$.
We show that there exist coordinates $a_\ell$ on $\cC$ such that the symplectic form on the total space is $L = \diff a_\ell \^ \diff z^\ell$ where $\ell$ labels the 1-dimensional torus factors of the fiber and $\diff z^\ell$ are their holomorphic 1-forms.
This follows from the defining properties of $L$, equivalent to the so-called SK integrability conditions. 
We show, furthermore, that the complex modulus of the $\ell$th torus factor is a function of $a_\ell$ alone, 
\begin{align}\label{c split t}
    Z^\ell(u) = Z^\ell(a_\ell) .
\end{align}
by leveraging the assumption that the SK geometry is completely split.
Finally, the scale symmetry of a scale-invariant CB implies that the $a_\ell$ have scaling weight 1 while the $Z^\ell$ have scaling weight 0, so the $Z^\ell$ must be locally constant on $\cC$.
A quick argument then shows that this is a basis independent statement, so we may conclude that $\t^{ij}$ --- which is generically related to $Z^{ij}$ through a basis change of the fiber --- is also locally constant on $\cC$.
Analytic continuation implies $\t^{ij}$ is therefore globally constant, and so the geometry is isotrivial.

A subtlety is that the definition of ``split'' that we use refers only to the abelian variety fibers as complex tori while ignoring their polarization (which is part of the definition of the SK geometry, as reviewed below).
In particular, although we typically express the SK structure in terms of a \emph{symplectic basis} with respect to the polarization given by the Dirac pairing, thus defining the complex modulus $\t^{ij}$, the 1-homology basis of the fiber in which \eqref{c split1} holds need not be a symplectic basis with respect to the polarization.
This means the $Z^{ij}$ modulus referred to above need not coincide with the $\t^{ij}$ modulus of the SK structure: even though $Z^{ij}$ is diagonal, there may be no EM duality frame in which $\t^{ij}$ is diagonal.
Nevertheless, the completely split property of $Z^{ij}$ together with scale invariance implies its isotriviality (local constancy), and the isotriviality of $Z^{ij}$ in turn implies the isotriviality of $\t^{ij}$.
The latter implication follows even if $\t^{ij}$ is not split simply because of the discreteness of the 1-homology basis change relating the split basis of $Z^{ij}$ to the symplectic basis of $\t^{ij}$.

If a geometry is split but not completely split, isotriviality no longer follows.
The simple counter example comes from the subset of split geometries that are direct product geometries containing a scale-invariant rank-$r{>}1$ factor that is not isotrivial.%
\footnote{Note that all scale-invariant rank-$1$ CB geometries are isotrivial, so the direct product of $r$ rank-$1$ geometries is completely split and indeed isotrivial.}
For instance, if there were a $2\times 2$ or larger block in a not completely split $Z^{ij}$, it could correspond to a non-isotrivial but scale-invariant factor of a direct product geometry.
In general, only fibers that split into all rank-$1$ factors are necessarily isotrivial.
This does not imply completely split geometries are direct product geometries, even though both are isotrivial, since completely split geometries may have some discrete ``twisting" of the local SK structures that renders the total fibration not a product.
It would be interesting to further explore split but not completely split geometries as they differ from product geometries only by some discrete global twist of the SK structure, and so may be amenable to a complete description in terms of ``twisting together" lower-rank geometries.

The completely split condition can be relaxed significantly while still maintaining isotriviality. 
The larger class of geometries, which we call completely \emph{iso-split}, are ones whose abelian varieties are \emph{isogenous} to a product of rank-1 factors rather than isomorphic.
(In the standard math terminology, reviewed in the appendix, the fibers of an iso-split geometry are \emph{decomposable} tori.)
One consequence of this strengthening of the isotriviality result is that all the rank-2 CB geometries whose generic fiber has an enhanced automorphism group (beyond the minimum $\Z_2$ group) are necessarily isotrivial.
This implies, for instance, that the only principally polarized rank 2 geometries which are not isotrivial are described by a smooth genus-2 Seiberg-Witten curve.  
This substantially simplifies the rank 2 classification program described in \cite{Argyres:2022lah}.

Throughout, we only work locally on the CB in an open, contractible, smooth subset $\cC$.
This ignores interesting aspects of the global geometry of completely split CBs, such as their singularities (around which the fibration has monodromies). 
An analysis of such global properties is relevant to the classification of completely split geometries as isotrivial geometries, and we leave these questions for future work to appear in \cite{ACDMMTW24}.

The rest of this note is organized as follows.
The next section briefly reviews the ingredients of local SK geometry, and the constraints on it that come from complex scale invariance.
In section \ref{section 3} we define the notion of split and completely split SK geometries, and then analyze the local SK geometry for a completely split geometry, showing that it is isotrivial if scale invariant, giving the details of the argument outlined above.
In section \ref{section 4} we generalize our results to the completely iso-split case, and describe some consequences of this result for the classification program of scale-invariant CB geometries.
An appendix reviews some pertinent facts about complex tori and abelian varieties.

\section{Local SK geometry}
\label{section 2}

We briefly review local SK geometry; some other reviews are \cite{Freed:1997dp, Argyres1998Lectures, Argyres1998Lectures2, Tachikawa:2013kta}.
This is the complex geometry of local patches of the CB component of the moduli space of vacua of 4d $\cN{=}2$ supersymmetric field theories.
Here ``local" means the geometry of open, contractible, and smooth subsets, $\cC$, of the CB component of the moduli space.
Note that we are deliberately working on \emph{contractible} local patches and will utilize this assumption at numerous points in the derivations that follow.
We will then specialize to CBs of SCFTs, whose SK geometries enjoy an additional complex scale symmetry action.

With respect to a choice of $\cN = 2$ supersymmetry algebra, the low-energy theory locally on the moduli space of $\cN \ge 2$ field theories consists of a product of massless free vector multiplets and massless neutral hypermultiplets.
Thus, the moduli space is locally a product, $\cC \times \cH$, of a local Higgs branch component $\cH$ parameterized by the vevs of the hypermultiplet scalars, and the local CB component $\cC$ parameterized by the vevs of the complex scalars in the vector multiplets. We are only concerned with the geometry of local patches $\cC$ on the CB where all vevs of the free hypermultiplet scalars are zero --- the so-called ``pure" component of the CB.

The bosonic part of the IR effective action of a rank-$r$ CB describes an $\cN = 2$ $\U(1)^r$ gauge theory,
\begin{align}\label{CB IREA}
    \cL_\cC &= \Im \left[ \t^{ij}(a) 
    \left(\del a_i \cdot \del \bar{a}_j 
    + \cF_i \cdot \cF_j \right) \right],
\end{align}
where $a_i$ are complex scalars of the vector multiplets, related by supersymmetry to the self-dual field strengths, $\cF_i$, of the $i$th $\U(1)$ gauge factor, $i=1,...,r$. 
$\t^{ij}$ is the symmetric complex matrix of $\U(1)^r$ gauge couplings to massive electrically and magnetically charged states on the CB.
Unitarity requires its imaginary part to be positive definite, $\cN{=}2$ supersymmetry requires that it depend only  holomorphically on the $a_i$ and furthermore that it satisfies the \emph{SK integrability condition}
\begin{align}\label{skcond}
    \del_a^{[i}\t^{j]k} = 0 ,
\end{align}
where $\del^i_a \doteq \del/\del a_i$. The
$a_i$ are commonly referred to as ``special coordinates'' on $\cC$.

The $\t^{ij}$ couplings are defined relative to a choice of normalization of the EM charges of massive states in the effective theory.
By defining the electric and magnetic charges of each $\U(1)$ gauge factor to be $e_i \deq \oint_{S^2}\star F_i \in \Z$ and $g_i \deq \frac{1}{2\pi} \oint_{S^2}F_i \in \Z$, respectively, a generic EM charge $(e,g)$ takes values in a rank-$2r$ lattice, $\L_{EM} \cong \Z^{2r}$.
The Dirac quantization condition defines the Dirac pairing  (a.k.a.\ Schwinger product), $J: \L_{EM} \times \L_{EM} \to \Z$, which is a physically observable, non-degenerate, integral skew pairing on the charge lattice \cite{Coleman:1985rnk}.

\paragraph{Local SK geometry as lagrangian fibration of abelian varieties.}

The IR effective action couplings $\t^{ij}(a)$ and the EM charge lattice $\L_{EM}$ together with its Dirac pairing $J$ define a \emph{polarized abelian variety} (p.a.v.), $(X_a,J)$, for each $a \in \cC$.
P.a.v.s are $r$-complex-dimensional algebraic tori, $X_a$, with a choice of discrete extra structure, $J$, the polarization, and are reviewed in appendix \ref{appA}.
The CB geometry thus comes with a fibration of rank-$r$ p.a.v.s $(X_a,J)$ over $\cC$ with a fixed polarization $J$.
As we will describe, once we restrict to a particular class of bases skew-diagonalizing the polarization $J$, the physical coupling matrix $\t^{ij}(a)$ appears in the description of $X_a$ as a complex torus through its period matrix. 
For each basis in this class, $\t^{ij}(a)$ takes values in the Siegel upper-half space of genus $r$, and the set of all such values forms an orbit under basis transformations between elements of this class.
These basis transformations correspond to EM-duality transformations of the CB IR effective action \eqref{CB IREA}.

The extra ingredients in the CB effective action --- the special coordinates and the integrability condition \eqref{skcond} --- give the {\it local SK geometry} on $\cC$.
This is conveniently encoded in the complex geometry of a $2r$-complex-dimensional space $\cX$ \cite{Donagi94, Seiberg:1994aj, Donagi:1995cf}, with the structure of a complex lagrangian fibration describing an algebraic integrable system.
$\cX$ is the total space of the holomorphic fibration $\pi: \cX \to \cC$ with p.a.v.\ fibers $X_u \doteq \pi^{-1}(u)$,
and is equipped with a holomorphic symplectic form, $L$, on $\cX$ such that the fibers are lagrangian, $L\vert_{X_u} = 0$.
Being a symplectic form, $L$ is a closed, non-degenerate $(2,0)$-form, and as we will review in this section, it implies the existence of special coordinates $a_i$ such that the SK integrability condition \eqref{skcond} holds.
(We will often refer to the closedness of $L$ as the SK integrability condition since it implies \eqref{skcond}).

Since $\cC$ is contractible, we can take coordinates $(x^a,u_j) \in \R^{2r}\times \C^r$ trivializing $\cX$ as a topological torus bundle.
Here $(u_j)$ are complex coordinates covering $\cC$ and $(x^a)$ are real periodic coordinates,
\begin{align}\label{x period}
    x^a \sim x^a +1, \quad a = 1, \ldots, 2r
\end{align}
on the fiber.
Note that the $u_j$ coordinates need not be special coordinates.
Then, fiber coordinates that are holomorphic with respect to the fibration $\pi$ can be taken to be complex linear combinations of the $x^a$, $z^j(u) = p^j_a(u) x^a + q^j(u)$, $j = 1, \ldots, r$, with $p^j_a$ and $q^j$ holomorphic functions on $\cC$, since there is always a local holomorphic section due to $\cC$ being contractible.
When the $z^j(u)$ are linearly independent, the possible additive shift, $q^j(u)$, can be set to zero by an appropriate $u$-dependent shift of the trivializing $x^a$ coordinates.     

The Dirac pairing defines a non-degenerate, integral, skew-symmetric bilinear form (i.e., a symplectic pairing), $J$, on the 1-homology lattice, $ H_1(X_u, \Z)$, of $X_u$. 
This lattice is identified with the lattice $\L_{EM}$ of charged massive states on the CB.
Being a symplectic lattice, there exists a basis $\{ \a_i, \b^i \}$ of $H_1(X_u,\Z)$ such that the matrix form of $J$ is given by
\begin{align}\label{J Delta}
    J = \bpm 0 & \D \\ -\D & 0 \epm \qq{with}  \D \doteq \text{diag} \{d_1, \ldots, d_r\}, 
    \quad d_i \in \N, 
    \quad d_i | d_{i+1}.
\end{align}
The $d_i$ are the \emph{invariant factors} of $J$, and we will call the resulting choice of $1$-homology basis a \emph{symplectic basis} for $J$, or \emph{$J$-symplectic basis}.

That $J$ defines a \emph{polarization} on $X_u$ means it satisfies certain compatibility conditions with the period matrix of $X_u$ --- the so-called Riemann conditions, reviewed in appendix \ref{appA}.
In particular, we can always specialize to a $J$-symplectic basis of $H_1(X_u, \Z)$ and a basis $\{\th^i\}$ of $H^{(1,0)}(X_u)$ such that the period matrix $\Pi$, thought of as a complex $r\times 2r$ matrix, takes the form
\begin{align}\label{tau Delta}
    \Pi \doteq \bigl( {\textstyle \int_\a \th, \int_\b \th} \bigr) 
    = \bigl(\D, \t \bigr) \qq{with} \t = \t^t, \qquad \Im \t > 0 ,
\end{align}
where $\D$ and the matrix form of $J$ satisfy \eqref{J Delta}.
Then, $\t$ is the $r\times r$ matrix of low energy $\U(1)^r$ gauge couplings defined in \eqref{CB IREA}.
We call the resulting pair of basis choices for the $1$-homology and $(1,0)$-cohomology a \emph{symplectic basis of} the p.a.v. $(X_u,J)$, or simply $X_u$ when the choice of polarization is understood.
When all $d_i =1$, the polarization is said to be \emph{principal}.

Different choices of polarization of $X_u$ give different CB geometries, and so, in particular, inequivalent low-energy effective actions \eqref{CB IREA} on $\cC$ with distinct $\U(1)^r$ gauge couplings. 
The $\{\a_i\}$ and $\{\b^i\}$ homology symplectic basis elements physically correspond to a choice of (normalized) ``electric" and ``magnetic" charges, respectively.
Since the $1$-homology basis and $J$ are discrete, they are constant on $\cC$.

Let $\{x^i, \, \hx_i \doteq x^{r+i}\}$ be real fiber coordinates whose corresponding dual $1$-forms $\{\diff x^a \}$ spanning $H^1(X_u)$ as a real vector space are dual to a $J$-symplectic $1$-homology basis, $\{\a_i, \b^i\}$, so that $\int_{\a_i} \diff x^j = \int_{\b^j} \diff\hx_i = \d_i^j$ and $\int_{\a_i} \diff\hx_j = \int_{\b^j} \diff x^i = 0$.
It is convenient to choose complex fiber coordinates
\begin{align}\label{z coords}
    z^i = \D^i_j x^j + \t^{ij}(u) \hx_j ,
\end{align}
which we write in matrix notation as $z=\D x + \t \hx$.
$(z,u)$ are holomorphic coordinates on $\cX$; the $u^i$ are good (i.e., single-valued, non-degenerate) coordinates on the base, but the $z$ are not good coordinates on the fiber as they suffer from periodic identifications by virtue of \eqref{x period}.
Let $\diff$ denote the total exterior derivative on $\cX$. 
In the $(z,u)$ coordinates, we decompose the exterior derivative into Dolbeault operators as $\diff = \del + \delb$, with $\del = \del_z \, \diff z + \del_u \, \diff u$.
Then,
the $1$-forms $\diff u$ and $\diff\bar u$ are globally defined on $\cX$, but $\diff z$ and $\diff \bar z$ are not because they depend on $\hx$, $\diff z = \D \diff x+\t \diff\hx +\diff\t \cdot \hx$.
Note that since $\t$ is a function of $u$ only, $\diff\t = \del_u^k\t \, \diff u_k$, where where $\del_u^k = {\del}/{\del u_k}$. 
Introduce the following 1-forms that are globally defined on $\cX$,
\begin{align}
    \th &\doteq \D \diff x + \t \diff\hx = \diff z - \diff\t \cdot \hx 
\end{align}
and their complex conjugates $\thb = \D \diff x + \bt \diff \hx$.
$\th$ is holomorphic, but not closed on $\cX$,
\begin{align}
    \del\th & = \diff\t \frac{1}{\t_-} (\th-\thb),&  
    \delb\th & = 0, &
    \t_- &\doteq \t - \bt .
\end{align}

In addition to the fibration structure, the total space has a holomorphic symplectic form, $L$, with respect to which the total space is a lagrangian fibration. 
This means that $L$ is a closed, non-degenerate $(2,0)$ form on $\cX$, and the restriction of $L$ to the fibers vanishes (so it contains no $\th^i \wedge \th^j$ terms).
$L$ is globally defined on all of $\cX$, so we can use the globally defined $1$-forms $du_i$ and $\th^i$ on $\cX$ to build it.
Thus, in terms of the $(z,u)$ coordinates, the most general $L$ is of the form
\begin{align}
    L &= \l_i^j \th^i\wedge \diff u_j + \m^{ij} \diff u_i\wedge \diff u_j ,
\end{align}
where $\l_i^j(z,\zb,u, \bar u)$ and $\m^{ij}(z,\zb,u, \bar u)$ are generic $\C$-valued functions on $\cX$.
$L$ being non-degenerate implies that $\det\l\neq0$.
$\l$ and $\m$ must be independent of $z$ and $\bar z$ since the fibers are compact.
$L$ is holomorphic so $\bar \del L = 0$, which implies that $\l$ and $\m$ are independent of $\bar u$, so $\l=\l(u)$ and $\m=\mu(u)$ are functions of $u$ alone.
Since $L$ is closed, $\diff L=\del L+\delb L=0$, implying $\del L =0$, which is
\begin{align}\label{L_closed}
    0 & = \del L = (\del_u^k \l^j_i) \diff u_k \wedge \th^i \wedge \diff u_j + \l^j_i(\del \th^i) \wedge \diff u_j + (\del_u^k \mu^{ij})\diff u_k \wedge \diff u_i \wedge \diff u_j  \\
    &= (\del_u^{[k} \l^{j]}_i) \D^i_n \diff u_k \wedge \diff x^n \wedge \diff u_j + \del^k_u(\l^{j]}_i \t^{in})\diff u_k \wedge \diff \hx_n \wedge \diff u_j + (\del_u^{[k} \mu^{i]j})\diff u_k \wedge \diff u_i \wedge \diff u_j. \nn
\end{align}
This is equivalent to the relations (using that $\D^i_n$ as defined in \eqref{J Delta} is always invertible), 
\begin{align}\label{L_closed_rels}
    \del_u^{[k} \l_i^{j]} & =0, &
    \l_i^{[k}\del_u^{l]}\t^{ij} &=0, &
    \del_u^{[k}\m^{i]j} &=0,
\end{align}
where we've inserted the first relation into the second term of \eqref{L_closed} to get the second relation.
The first relation implies $\l_i^j \diff u_j$ for each $i$ is a closed $1$-form in $H^{(1,0)}(\cC)$.
$\cC$ is contractible by assumption, so this cohomology is trivial, implying each $\l_i^j$ is $\del_u$-exact,
\begin{equation}\label{L eq1}
    \l_i^j(u) = \del_u^j a_i(u) , 
\end{equation}
where the $a_i(u)$ are some holomorphic functions on $\cC$.
The $a_i$ are singled-valued and non-degenerate coordinates on $\cC$ since $\det \l\neq0$.
Similar reasoning applied to the third relation implies that $\m^{ij}(u)=\del_u^{[i} b^{j]}(u)$, with $b^j (u)$ some holomorphic functions on $\cC$.

The second relation in \eqref{L_closed_rels} implies the SK integrability condition \eqref{skcond} since, from \eqref{L eq1} we have $\del_u^l = (\del_u^l a_m) \del_a^m =\l^l_m \del_a^m$, so the second relation in \eqref{L_closed_rels} can be rewritten as $\l^{[k}_{[i} \l^{l]}_{m]} \del_a^{[m}\t^{i]j} = 0.$
Because $L$ is non-degenerate, $\l^i_j(u)$ is invertible in all of $\cC$, implying the SK integrability condition $\del_a^{[m} \t^{i]j} = 0$.
From this it follows that the $a_i$ are the special coordinates used to define the low-energy effective action in \eqref{CB IREA}.

Define the holomorphic $(1,0)$-form $b(u) = b^j \diff u_j$ on $\cC$.
Then
\begin{align}\label{L2}
    L & = \w^j \^ \diff a_j, & 
    \w^j & \doteq \th^j + \del_a^j b .
\end{align}
The $\{\w^j\}$ form a set of holomorphic (but not closed) $(1,0)$-forms on $\cX$.
Upon restriction to the fiber $X_u$, $\w^j|_u = \th^j|_u = \D \diff x + \t(u) \diff \hx$, and so form a basis of the fiber $(1,0)$-cohomology whose fiber periods relative to the $J$-symplectic basis $\{\a_i, \b^j\}$ satisfy \eqref{tau Delta}.
Hence, locally, the differentials of the (dual) special coordinates on the CB are determined by the fiber periods of $L$ up to constants,
\begin{align}\label{sp coords}
    \int_{\a_i} L &=  \diff a_j \int_{\a_i} \w^j =  \D^j_i \diff a_j  , & \int_{\b^i} L =  \diff a_j \int_{\b^i} \w^j =  \t^{ij} \diff a_j .&
\end{align}
Note that the fiber periods of $L$ are independent of its $\diff b = \mu^{ij} \diff u_i \wedge \diff u_j$ term.

\paragraph{Scale invariance.}

The CB of a SCFT enjoys a holomorphic $\C^\times$-action as a consequence of the spontaneously broken dilatation and $\U(1)_R$ symmetries on the CB.
For $\l\in\C^\times$ and $u$ any (smooth or non-smooth) point of the CB, we denote this action by $\l: u \mapsto \l\circ u$.
By the assumed uniqueness of the conformal vacuum, only one point of the CB is fixed by all of $\C^\times$, namely, the conformal vacuum where the where dilatation and $\U(1)_R$ transformations are preserved.
Define $B_\e^* \doteq \{ \l {\in} \C \ \text{with}\ |\l-1|<\e \ \text{and}\ \l\neq1 \}$ to be a small punctured disk of elements $B_\e^* \subset \C^\times$ close to the identity element.
Therefore, for $\e$ small enough, all $\l\in B_\e^*$ will act on our smooth contractible subset $\cC \subset$ CB holomorphically and transitively.
Though some points in $B_\e\circ \cC$ may be outside of $\cC$, by restricting to  a sufficiently small open subset $\cC_\d \subset \cC$, we can ensure that $B_\e\circ \cC_\d \subset \cC$.
This follows readily from the the Picard–Lindel\"of theorem which shows that the local flow of $\l_t(z) \circ u$ about $u \in \C$ and for $\l_t(z) \deq \exp(t z)$ for some $|z| < \log(1+\e)$ and $t$ in a small enough interval containing 0, has a unique solution that resides in an open neighborhood of $u$ that is contained in $\cC$.
Thus on a small enough smooth, simply connected local patch $\cC_\d$, by the domain-straightening theorem for vector field flows we can choose complex coordinates $(u_1, \ldots, u_r) \in \C^r$ which diagonalize the $B_\e$ action,
\begin{align}
    B_\e : \qquad  u_k & \mapsto \l\circ u_k \doteq \l^{\D_k} u_k , & 
    \l & \in B_\e , 
\end{align}
where the scaling weights, $\D_k$, can be chosen to be real.%
\footnote{Note that the set of scaling weights of local coordinates is not uniquely defined, but may depend on the choice of coordinates.}

The $\C^\times$ scaling action leaves the Dirac pairing and $\t_{ij}$ effective couplings unchanged, and so acts as an automorphism of the polarized abelian variety fibers.
Furthermore, the symplectic form $L$ transforms homogeneously with weight $\D_L=1$, since its periods compute masses (which are defined to have weight one).
Since the fiber periods of $L$ are the special coordinates $a_i$ and their duals, the special coordinates also have mass dimension 1, $\D_{a} =1$.
Taking $\l\in B_\e$, it follows that we can take 
\begin{align}\label{scaling symm}
    \pi^{-1}(\l\circ u) & = \pi^{-1}(u) , &
    B_\e : L & \mapsto \l L ,
\end{align}
for $u\in \cC_\d$.
The integer periodicity \eqref{x period} of our chosen fiber coordinates $x^a$ implies they have scaling weight zero.
Coupled with the expression \eqref{L2}, it follows that the $\w^{j}$ and $\th^i$ have zero scaling dimension, $\D_{\w}= \D_{\th} =0$.
From the definition \eqref{z coords} of the complex $z^i$ fiber coordinates, it also follows that $\D_{z} = \D_{\t}=0$.
We summarize the scaling weights of these various quantities,
\begin{align}\label{scale1}
    \D_z = \D_\t = \D_\th  = \D_\w  & = 0, & \D_a & = 1 .
\end{align}

\section{Completely split scale-invariant SK geometries are isotrivial}\label{section 3}

``Completely split'' rank-$r$ SK geometries are ones whose generic abelian variety fiber, considered as a complex torus (i.e., without regard to its polarization), is a direct product of $r$ genus-$1$ Riemann surfaces.
We solve the SK integrability conditions for a completely split geometry on open, contractible, smooth subsets, $\cC$, of a scale-invariant CB, and show that all solutions are isotrivial.

\subsection{Split SK geometries}

A split complex torus is one which is isomorphic to the direct product of two or more lower-dimensional complex tori.
We define split complex tori and abelian varieties more carefully in appendix \ref{appA}, where we emphasize the different notions of isomorphism for complex tori and p.a.v.s.

A rank-$r$ \emph{split SK geometry} is a lagrangian fibration $\pi: \cX \to \cC$ whose generic fibers, $X_u$ for $u\in \cC$, are isomorphic (as complex tori) to the direct product of $n$ lower-rank complex tori, $X^\ell_u$, of (complex) dimensions $r_\ell$ for $\ell = 1, \cdots, n$.
For concreteness, we will consider the $n=2$ case in which the generic fiber splits into two complex tori, $X_u \simeq_\C X_u^1 \times X_u^2$, of ranks $r_1$ and $r_2$ satisfying $r = r_1 + r_2$.
This means that there is a basis of the fiber 1-homology,
$\{ \ta_i, \tb^i\}$, and of the fiber $(1,0)$-cohomology, $\{ \tth^i \}$, such that
\begin{align}\label{split_basis}
    \int_{\ta}\tth &= \mathds 1_{r } , 
    & \int_{\tb}\tth &\deq Z, &
    &\text{with} & 
    Z(u) &\deq \bpm Z^1(u) & \\ & Z^2(u) \epm ,
\end{align}
where each $Z^\ell$ is an $r_\ell \times r_\ell$ matrix. 
(We are using a matrix notation in which $\ta$ and $\tb$ are row matrices and $\tth$ is a column matrix.)
We call such a basis a \emph{split} basis for $X_u$.
Note that $Z$ is not necessarily symmetric and positive definite, and does not generically correspond to the physical coupling of the low-energy theory on the CB.
Furthermore, note that the Dirac pairing, $J$, will generally not be in the symplectic form $J = \e \otimes \D$ with respect to the $1$-homology basis $\{ \til \a_i, \til \b^i \}$, c.f. \eqref{J Delta}. 

Being split is a local condition, and over a smooth local patch $\cC$ it implies that the total space is a fibered product, $\cX \cong \cX^1 \times_\cC \cX^2$, where $\cX^\ell$ is the total space of a $(r + r_\ell)$-complex dimensional torus bundle  over $\cC$ with generic fiber a $r_\ell$-$\dim_{\C}$ torus $X^\ell_u$.
Note that since this is a fibered product and not a cartesian product of bundles, this does not imply that the CB geometry is globally the direct product of rank-$r_\ell$ CB geometries.

Although $X^\ell_u$ is canonically a rank-$r_\ell$ abelian variety (since it is a sub-torus of $X_u$), we are not equipping it with a polarization and will only be concerned with its structure as a complex torus. 
In fact, being split is a condition on the generic fiber as a complex torus and not as a polarized abelian variety.
Since $X_u$ is a polarized abelian variety, there exists a symplectic 1-homology basis $\{\a_i,\b^i\}$ and $(1,0)$-cohomology basis $\{\th^i\}$ such that \eqref{tau Delta} holds.
Define $R \in \GL(2r, \Z)$ and $G\in \GL(r,\C)$ by
\begin{align}\label{symp to split}
    \bigl( \a \ \b \bigr) &\doteq 
    \bigl( \ta \ \tb \bigr) R^{-1}, &
    \th & \doteq G^{-t}\, \tth .
\end{align}
In going from the split (with tildes) basis \eqref{split_basis} to the symplectic (without tildes) basis \eqref{tau Delta}, the period matrices $\til\Pi=(\mathds 1_r, Z)$ and $\Pi=(\D,\t)$ are related as $\Pi(u) = G \til\Pi(u) R^{-1}$.
Writing $R^{-t} \doteq \bspm d & c \\ b & a \espm$ in $r\times r$ block form, this implies 
\begin{align}\label{t to tt}
   \t(u) &= (a Z^t(u) + b) (c Z^t(u) + d)^{-1} \D^t,
\end{align}
where we applied \eqref{PM_basis_trans} and $\t=\t^t$ that results from the Riemann conditions.
If $X_u$ is split, it need not have a symplectic basis which is simultaneously a split basis, i.e., one for which the Dirac pairing takes the form \eqref{J Delta} while $\int_\a \th = \D$ and $\int_\b \th = \t(u) = \t^{1}(u) \oplus \t^2(u)$ with each $\t^\ell$ being symmetric with positive definite imaginary part.
Equivalently, the change of lattice basis $R^{-t}$ need not be in $\Sp_J(2r,\Z)$ where $J$ takes the matrix form $\e \otimes \D$.
If it were, this matrix form of $J$ would be preserved in the split basis and the transformation \eqref{symp to split} would be an isomorphism of p.a.v.s; see appendix \ref{appA}.
It is in this sense that being split is a property of $X_u$ as a complex torus and not as a polarized abelian variety.

We say an SK geometry is \emph{completely split} if the generic fiber is a product of $r$ rank-1 tori, i.e., if there is a \emph{completely split} basis of the $1$-homology and $(1,0)$-cohomology such that $\int_\ta \tth = \mathds 1_{r }$ and $\int_\tb \tth = \text{diag}(Z^1, \cdots, Z^r)$ is diagonal as in \eqref{c split1}, so
\begin{align}\label{c split2}
    X_u \cong_{\C} X^1_u \times \cdots \times X^r_u.
\end{align}
Thus, each $X^\ell_u$ factor is a rank-1 complex torus, i.e., a genus-one Riemann surface which can vary holomorphically over $\cC$.
Note that at rank $r=2$ a split geometry is necessarily completely split, and there is only a distinction between split and completely split starting at ranks $r\ge 3$.

An SK geometry is \emph{isotrivial} if $\t$ is locally constant on $\cC$.
Isotriviality is really a property of $\cX$ as a fibration of complex tori, rather than of abelian varieties, since, by \eqref{t to tt} and its inverse, $Z$ is locally constant if and only if $\t$ is.
This means that if one shows the period matrix $\Pi(u) = \Pi_0$ describing each abelian variety fiber $X_u$ is constant on $\cC$, then it follows that the p.a.v.\ fibers $(X_u \cong_{\C} X_0, J)$ with polarization $J$ are all also isomorphic (as p.a.v.s) since they are equipped with the same polarization which is locally constant on $\cC$.
For our purposes, it will thus suffice for us to show that $Z$ is constant on $\cC$ to deduce that $\cX$ is isotrivial without reference to the polarization that each abelian variety fiber is equipped with.

\subsection{Completely split SK integrability conditions}

To solve the SK integrability conditions for a completely split geometry, we follow the same approach that was taken in section 2 with the modification that we will choose to work in the split basis of the fiber as opposed to a symplectic basis.
We denote the split basis coordinates and cycles with tildes.
In particular, trivialize the fibration $\cX$ on $\cC$ as a topological torus bundle using local coordinates $(\til x^a, u_j) \in (\R/\Z)^{2r} \times \C^r$ as in the discussion around \eqref{x period}, and choose a fiber 1-homology basis $\{\ta_i,\tb^i\}$ dual to the $\{\tx^i, \, \htx_i \doteq \tx^{r+i}\}$ real periodic coordinates, so that $\int_{\k_i} \diff \tx^j = \int_{\tb^j} \diff\htx_i = \d_i^j$ and $\int_{\ta_i} \diff\htx_j = \int_{\tb^j} \diff \tx^i = 0$.
Since, by assumption, the fibers are completely split, there are complex fiber coordinates
\begin{align}\label{split z coords}
    \tz \doteq \tx + Z(u) \, \htx ,
\end{align}
(written in matrix notation) such that the complex period matrix takes the form
\begin{align} \label{com split basis} 
   Z(u) = \bpm \z^1(u) & & & \\ & \z^2(u) & & & \\ & & \ddots & & \\ & & & \z^r(u) \epm,
\end{align}
as in \eqref{split_ct_2}.
This split basis of $X_u$ as a complex torus is the analogue of \eqref{z coords} which gave a symplectic basis of $X_u$ as a p.a.v.

As before, we introduce a corresponding basis of globally defined holomorphic 1-forms on $\cX$ by
\begin{align}
    & \diff u & 
    &\text{and}&
    \tth &\doteq \diff \tx + Z \, \diff\htx = \diff \tz - \diff Z \cdot \htx ,
\end{align}
in terms of which the most general lagrangian $(2,0)$-form on $\cX$ can be written in $(z,u)$ coordinates as
\begin{align} \label{SKI split 0} 
   L = \tl^j_{\ i} \tth^i \wedge \diff u_j + \tm^{ij} \diff u_i \wedge \diff u_j,
\end{align}
in terms of some complex coefficient functions $\tl^j_{\ i}$ and $\tm^{ij}$ on $\cX$.
Just as in the discussion around \eqref{L_closed}, non-degeneracy of $L$ implies $\det \tl \neq 0$, holomorphy of $L$ implies $\tl$ and $\tm$ are holomorphic, and compactness of the fibers imply they are independent of $\tz$.  
Then closedness of $L$ gives the same conditions \eqref{L_closed_rels}, but with tildes on $\l$ and $\m$ and with $\t$ replaced by $Z$, giving the solutions
\begin{align}\label{split SKI} 
   \tl^j_{\ i} &= \del_u^j {\til a}_i, &
   \tm^{ij} &= \del_u^{[i} {\til b}^{j]}, &
   \del_{\til a}^{[i} Z^{j] k} &= 0 ,
\end{align}
where $\til a_i$ are some independent holomorphic functions, which we will refer to as the `split coordinates' on our contractible patch $\cC$, and ${\til b}^j$ are some holomorphic functions on $\cC$.

The last condition in \eqref{split SKI} is the analog of the SK integrability condition \eqref{skcond}, and implies the existence of holomorphic functions $\fz^k(\til a)$ on $\cC$ such that $Z^{jk} = \del_{\til a}^j \fz^k$.
Now the completely split assumption, $Z^{jk} = \z^j \d^{jk}$ (no sum on $j$), implies
\begin{align}\label{} 
    \del_{\til a}^j \fz^k &= 0 \qquad 
    \text{for all} \qquad j \neq k.
\end{align}
Thus $\fz^k(\til a) = \fz^k(\til a_k)$ is a function only of the $\til a_k$ coordinate on $\cC$.
So, since $\z^k = \del^k_{\til a} \fz^k$, the complex modulus of the $k$th  rank-$1$ torus factor of the split fiber depends only on the $\til a_k$ coordinate on $\cC$,
\begin{align}\label{Z split scale} 
   Z^{ij}(\til a) = \diag\bigl( \z^1(\til a_1), \dots, \z^r(\til a_r) \bigr).
\end{align}

To summarize, we have found that the most general, local form of a completely split SK geometry on a local patch $\cC$ of the CB can be written with respect to coordinates $u \in \cC$ and the split basis of the fiber \eqref{com split basis} such that the holomorphic symplectic form $L$ takes the form
\begin{align} \label{Csplit L} 
L = \tw^j \wedge \diff {\til a}_j
\end{align}
where $\{\til a_i(u) \}$ are `good' holomorphic coordinates on $\cC$, i.e. $\det(\del_u^k \til a_j) \neq 0$, and we've defined the global, holomorphic $(1,0)$-forms $\tw^j$ on $\cX$ by
\begin{align} \label{Csplit 1f} 
\tw^j \deq \tth^j - \del_{\til a}^j \tm^k \diff u_k .
\end{align}
And $L$ is closed only if the complex modulus $\z^j(g_j)$ of the $j$th rank-1 torus factor of the split fiber depends only on the $j$th coordinate component ${\til a}_j(u)$.

\paragraph{Scale invariance.}

By the same discussion of the complex scaling action on $\cC$ given in section \ref{section 2}, it follows that we have scaling weights 
\begin{align}\label{scale2}
    \D_{\tz} &=
    \D_{Z} = \D_{\tth} = \D_{\tw} = 0, &
    \D_{\til a} &= 1 ,
\end{align}
just as in \eqref{scale1}.
In particular, the $\til a_k(u)$ are weighted homogeneous of weight one, and $\D_Z=0$, so the scaling symmetry acts as
\begin{align}\label{a scaling}
    \til a_k ( \l\circ u ) &= \l a_k(u) , &
    \z^k(\l\circ u) &= \z^k(u) ,&
    \l \in B_\e ,& & u\in \cC_\d .
\end{align}
Since by \eqref{Z split scale} $\z^k = \z^k(\til a_k)$, it follows that $\l \circ \z^k(\til a_k) = \z^k(\l \til a_k)$, but to match the scaling weights \eqref{scale2} we must have
\begin{align}
    \z^k(\l \til a_k) = \z^k(\til a_k),
\end{align}
which means the $\z^k$ have to be constant functions on $\cC_\d$,
\begin{align}\label{c scaling}
    \z^k(\til a_k) &= \text{constant},
\end{align}
and so the completely split SK geometry is isotrivial.
This is because since it is locally constant on the CB, it is constant on the whole CB by analytic continuation.

\subsection{Examples of completely split rank-2 geometries}

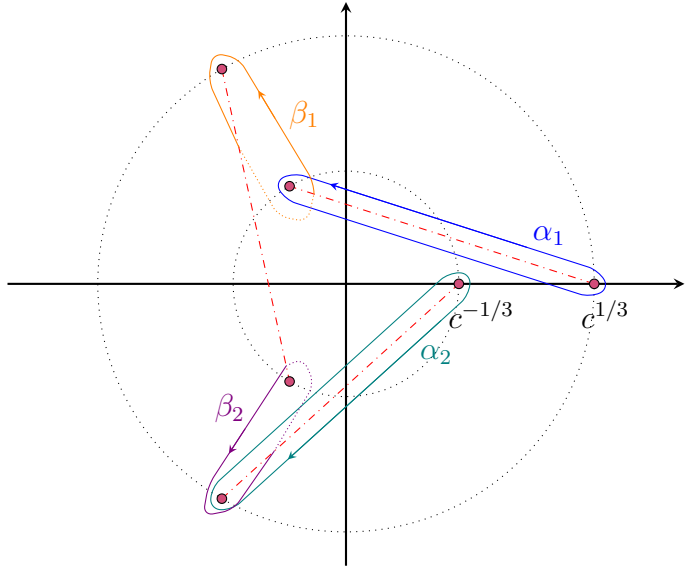
\begin{figure}[t]
    \centering
    \begin{tikzpicture}[scale=1.5]
        \draw[dotted] (0,0) circle (1);
        \draw[dotted] (0,0) circle (2.2);
        \draw[thick,stealth-]  (0,2.5) -- (0,-2.5);
        \draw[thick, -stealth] (-3,0) --  (3,0);
        \draw[color=red,dashdotted] (2.2,0) -- (-0.5,0.866025);
        \draw[color=red,dashdotted] (-1.1,1.90526) -- (-0.5,-0.866025);
        \draw[color=red,dashdotted] (-1.1,-1.90526) -- (1,0);
        \draw[fill=purple!70] (1,0) circle (1.2pt);
        \draw[fill=purple!70] (-0.5,0.866025) circle (1.2pt);
        \draw[fill=purple!70] (-1.1,1.90526) circle (1.2pt);
        \draw[fill=purple!70] (-0.5,-0.866025) circle (1.2pt);
        \draw[fill=purple!70] (-1.1,-1.90526) circle (1.2pt);
        \draw[fill=purple!70] (2.2,0) circle (1.2pt);
        \node[below] at (1.2,-0.05){$c^{-1/3}$};
        \node[below] at (2.3,-0.05){$c^{1/3}$};
        \draw[color=blue, rounded corners=4pt] (2.23849, 0.12) -- (-0.46151, 0.986025)--(-0.635, 0.909326) --(-0.53849, 0.746025) -- (2.16151, -0.12)--(2.335, -0.0433013)--cycle;
        \draw[color=blue, -stealth,shorten >=0.5cm,shorten <=1cm] (2.23849, 0.12) -- (-0.46151, 0.986025);
        \draw[color=teal, rounded corners] (0.909273, 0.1) -- (-1.19073, -1.80526)--(-1.205, -2.00052)--(-1.00927, -2.00526)--(1.09073, -0.1) -- (1.105, 0.095263) -- cycle;
        \draw[color=teal, stealth-,shorten >=0.5cm,shorten <=1cm] (-1.00927, -2.00526)--(1.09073, -0.1);
        \draw[color=orange,rounded corners] (-0.284, 0.796743) -- (-0.284, 0.896743) -- (-0.95,2) -- (-1.18,2.05)-- (-1.22,1.8) -- (-0.95, 1.21244);
        \draw[color=orange, -stealth,shorten >=0.5cm,shorten <=1cm] (-0.284, 0.896743) -- (-0.95,2); 
        \draw[color=orange, densely dotted, rounded corners] (-0.95, 1.21244)-- (-0.6, 0.6)-- (-0.34,0.55) --(-0.284, 0.796743);
        \draw[color=violet, rounded corners] (-0.53, -0.727461) -- (-1.22,-1.78526) -- (-1.268, -2.05768) -- (-0.99927, -2.00526) --(-0.638, -1.4861);
        \draw[color=violet, -stealth, shorten >=0.5cm,shorten <=1cm](-0.53, -0.727461) -- (-1.22,-1.78526);
        \draw[color=violet, rounded corners, densely dotted](-0.638, -1.4861)-- (-0.28,-0.886025) -- ({-0.5*0.76},{-0.866025*0.76}) -- (-0.53, -0.727461);
        \node[above,color=blue] at (1.8,0.25){$\alpha_1$};
        \node[below,color=teal] at (0.8,-0.45){$\alpha_2$};
        \node[right,color=orange] at (-0.6,1.5){$\beta_1$};
        \node[left,color=violet] at (-0.8,-1.1){$\beta_2$};
    \end{tikzpicture}
    \caption{The (canonical) basis of $H_1(X^c[D_6],\Z)$ used in \cite{RN_split} in order to calculate the period matrix in equation \eqref{eq:d6}.}
    \label{fig:paths}
\end{figure}

\begin{example}\label{ex:d6} 

{\bf Jacobian varieties which split.}

\medskip
\noindent Coulomb branch SK geometries with principal polarization are often described in terms of Jacobian varieties of families of smooth genus-$r$ Riemann surfaces --- the Seiberg-Witten curves of the geometry.
While such Jacobian varieties do not split as p.a.v.s, they can split as complex tori.

For example, consider the family of genus-2 Riemann surfaces with automorphism group $D_6$. 
They can be described by the hyperelliptic curve
\begin{gather}\label{eq:su3}
    X^c[D_6] = \{ (x,y)\in\C^2:\, y^2=(x^3-c)(x^3-c^{-1})\},
\end{gather}
where $c$ is an arbitrary complex parameter.  (When $c=\pm1$ the curve degenerates by pinching two non-intersecting cycles, giving a sphere with two pairs of points identified; when $c=0, \infty$ the curve degenerates by pinching a homologically trivial cycle, giving a bouquet of two tori; when $c=\pm i$ the curve remains smooth but its automorphism group is enhanced to $\Z_3 \rtimes D_4$;
and finally, when $c=(3\sqrt3 \pm 5)/\sqrt{2}$ the curve is smooth but its automorphism group is enhanced to $\GL_2(3)$.)
Using the basis of $H_1(X^c[D_6],\Z)$ shown in figure \ref{fig:paths}, and by demanding automorphism invariance, it can be shown that for an appropriate basis of (1,0)-forms the period matrix of $X^c[D_6]$ is given by \cite{RN_split}
\begin{gather}\label{eq:d6}
    (\Delta \ Z) = \begin{pmatrix}
        1 & 0 & 2z & z \\
        0 & 1 & z & 2z
    \end{pmatrix}.
\end{gather}
Here we see that $Z$ is not diagonal, so $J(X^c[D_6])$ does not obviously split as a p.a.v..  
(In fact, as we review in appendix \ref{app Jac}, since it is a Jacobian variety of a smooth Riemann surface, it does not split as a p.a.v.)
However, we can now perform a basis change of $1$-cycles and $(1,0)$-forms to get
\begin{gather}
    G \cdot (\Delta \ Z)\cdot R^{-1} = \begin{pmatrix}
        1 & 0 & 3z & 0 \\
        0 &  1 & 0 & z
    \end{pmatrix},
\end{gather}
where
\begin{gather}
    G =  \begin{pmatrix}
        \ph{-}2 & -1 \\
        -1 & \ph{-}1
    \end{pmatrix}, \qquad
    R= \begin{pmatrix}
        \ph{-}2 & -1 & \ph{-}0 & \ph{-}0\\
        -1 & \ph{-}1 & \ph{-}0 & \ph{-}0\\
        \ph{-}0 & \ph{-}0 & \ph{-}1 & \ph{-}0\\
        \ph{-}0 & \ph{-}0 & -1 & \ph{-}1
    \end{pmatrix}.
\end{gather}
While this is a perfectly valid change of basis ($\det G \neq 0$ and $\det R=1$), it is not a symplectic basis change since $R\notin \Sp(4,\Z)$. 
As such, this induces an isomorphism 
\begin{gather}
    J(X^c[D_6]) \cong E_{3z} \times E_z, \quad\quad 
    E_\t \doteq \C/(\Z+\t\Z),
\end{gather}
only at the level of complex tori. 
Note that $J(X^a[D_6])$ and $E_{3z} \times E_z$ are both principally p.a.v.s, but they are not isomorphic as p.a.v.s.

Since the whole family of curves \eqref{eq:su3} split, it follows from our result that a scale-invariant CB geometry with Seiberg-Witten curve of the form \eqref{eq:su3} must be isotrivial.  
This means that the parameter $c$ in the curve (or, equivalently, the parameter $z$ introduced in \eqref{eq:d6}) must be constant over the CB, and so will have the interpretation of an exactly marginal coupling of the associated CFT.
Indeed, precisely 2 such Seiberg-Witten curves were found in \cite{Argyres:2022fwy, Argyres:2023eij}, corresponding to to the $\cN=4$ $\su(3)$ and $G_2$ SYM theories. 
As these theories enjoy $\cN=4$ supersymmetry, they must be isotrivial. 
This is consistent with our derivation above, as the fiber is generically split.

It is not difficult to find other examples of splitting Jacobians in genus 3 and 4. For example, the Jacobians of Klein's quartic \cite{Rauch71} in genus 3 and Bring's curve in genus 4 \cite{brings_split} also split as complex tori. 
\end{example}

\begin{example}

{\bf Split complex tori which admit distinct polarizations.} 

\medskip
\noindent Throughout our discussion we have stressed that only splitness at the level of complex tori is required for isotriviality, as opposed to the reducibility of the polarized abelian variety in question. 
Let us provide an example where this polarization independence is manifest, polarizing a split complex torus in two different ways.

Consider the split complex torus $X= E_{\r}\times E_{\r}$ with $\r=e^{\pi i/3}$. 
There are several ways to polarize $X$--- the most obvious one being the principal polarization $\e\oplus \e$ inherited from the principal polarization $\e$ on each torus factor. 
By definition, this makes $(X,\e\oplus \e)$ a reducible abelian variety with $\mathsf{Aut}(X,\e\oplus \e) = \Z_6^2\rtimes \Z_2\cong G(6,1,2)$. 
Here $G(6,1,2)$ refers to the Shephard Todd classification of complex reflection groups \cite{ST54}.
Since $G(6,1,2)$ has degrees $\{6,12\}$ and reflections of order 2 and 6, one concludes that the CFT corresponding to the orbifold CB $\C^2/G(6,1,2)$ is the isotrivial rank-2 $E_8$ Minahan-Nemeschansky theory (entry 1 in table 1 of \cite{Martone:2021ixp}). 
    
On the other hand, we can polarize $X$ in a non-principal way. 
In fact, there is a polarization with invariant factors 1 and 6 that enjoys a $G_5$ automorphism group \cite{Fujiki88}, where again $G_5$ refers to the Shephard-Todd classification.
$G_5$ also has degrees $\{6,12\}$, but only has reflections of order 3.
The (relative)%
\footnote{\label{relQFT}In the context of $\cN=2$ SCFTs, being a \emph{relative} CFT means the SK geometry describing its CB is not principally polarized and therefore its homology lattice cannot be associated to a charge lattice of (possibly) probe line operators that contains the physical charge lattice of finite energy BPS states as a (proper) sublattice.
See \cite{Argyres:2022fwy} for more details on this interpretation of the SK geometry associated to a CB.}
CFT corresponding to the CB orbifold (and therefore isotrivial) geometry $\C^2/G_5$ was discussed in \cite{Cecotti:2021ouq}, where it was conjectured to possess $\cN=3$ supersymmetry.%
\footnote{We thank S. Cecotti for pointing out this example.}
\end{example}


\section{Completely iso-split scale-invariant CBs are isotrivial}
\label{section 4}

In this section, we ask if it's possible to weaken the assumption that the generic fiber of a scale-invariant SK geometry is isomorphic to a direct product of rank-1 tori, i.e., that it completely splits, while still maintaining isotriviality.
In fact, there is a natural class of maps between complex tori that can achieve this: such mappings are called \emph{isogenies}.
As reviewed in appendix \ref{appA}, isogeny is a weaker notion of equivalence between complex tori which amounts to the lattice defining one torus being mapped to a finite-index sublattice of the other, or, equivalently, an isogeny map is a finite covering of one complex torus by another.
The discreteness of lattice inclusions gives isogenies a ``rigidity" which makes them ``almost" isomorphisms.

Replacing isomorphism with isogeny in the definition of a split complex torus then leads to the definition of an \emph{iso-split} (scale invariant) SK geometry, which is defined via the pullback of the dual isogeny and possesses a generic fiber that is isogenous to a product of lower rank complex tori, i.e. is decomposable.
We also demand that the isogeny be continuous as a function of $u \in \cC$.
Then, the kernel of the isogeny 
as a finite subgroup of the torus is constant over $\cC$, since it is discrete.

When all the lower-rank torus fibers are rank $1$, we get a \emph{completely iso-split} SK geometry.
In this section, we show that scale-invariant SK geometries in which the generic fiber is completely iso-split are also isotrivial.
We do this by constructing an isogenous scale invariant SK geometry for which we can apply the results of section \ref{section 3} to conclude it is an isotrivial geometry; this immediately implies the non-completely split SK geometry is isotrivial after appealing to the rigidity of isogenies.
At the end of this section we will describe some implications for the classification program of scale-invariant SK geometries that results from this strengthening of our isotriviality result.

As we have done throughout this note, it is understood that we are considering a holomorphic fibration $\pi: \cX \to \cC$ with a base space $\cC$ that is a smooth, contractible, open, and scale-invariant patch of the total CB moduli space, and whose generic fiber is a p.a.v.\ $(X_u, J)$.
We now assume that the complex torus of the generic fiber, $X_u \cong \pi^{-1}(u)$, $u\in\cC$, is completely iso-split, so there is an isogeny $f_u : X_u \to \til X_u$ that maps the (non-completely split) complex torus $X_u$ fiber to a completely split complex torus 
\begin{align}\label{iso-fibers}
    \til X_u = \til X_u^1 \times \cdots \times \til X_u^r ,
\end{align}
where each $\til X_u^i$ is a rank-1 complex torus.
This then defines a holomorphic fibration $\til\pi: \til\cX \to \cC$ with fibers $\til \pi^{-1}(u) \cong \til X_u$, which we'll call the isogenous fibration.

As we describe in appendix \ref{appA}, $\til X_u$ is an abelian variety if $X_u$ is.  
Furthermore, one can check that the holomorphic symplectic form $L$ on $\cX$ defining the SK structure is pulled back by the dual isogeny to a holomorphic symplectic form $\til L$ on $\til \cX$, giving the isogenous fibration the structure of an SK geometry.
Finally, if $\cX$ is scale invariant --- i.e., has a free holomorphic $\C^\times$ symmetry action preserving the fibers as in \eqref{scaling symm} --- then the isogenous fibration $\til\cX$ is also scale invariant, since the $\C^\times$ action pulled back by the dual isogeny necessarily preserves the fibers $\til X_u$.

More generally, this discussion shows that CB geometries fall into equivalence classes under isogeny.
In particular, if they have isogenous fibrations and their SK structures on $\cC$ are related by the isogeny, we will call them \emph{locally isogenous CB geometries}.
(``Locally" because this argument only applies for open, smooth, contractible subsets, $\cC$, of the CB.)
An immediate consequence of this result is that, starting with a completely split SK geometry $\cX$ on a local patch $\cC$, one can define (multiple) SK geometries $\til \cX_i$ on $\cC$ which are also isotrivial simply by deducing the (inequivalent) isogenies $f_i: X \to \til X_i$ from the completely split complex torus fiber $X$ of $\cX$ to the complex tori $\til X_i$ of $\til \cX_i$ --- the latter complex tori/SK geometries are only guaranteed to be completely decomposable/iso-split.
Note that although it follows that the SK geometry of $\til \cX_i$ on the total CB is isotrivial, the isogeny $f_i$ only relates the local SK geometries of $\cX$ with $\til \cX_i$ on $\cC$.
This inequivalence of the global CBs under isogeny follows because non-trivial isogenies do not preserve the automorphism group of the p.a.v.\ fiber and because the EM monodromy group on the CB is a subgroup of the automorphism group of the fiber in isotrivial geometries.
Furthermore, as discussed in the appendix, non-trivial isogenies also do not preserve the polarization of the fibers, so locally isogenous CB geometries presumably correspond to different relative versions of a given (absolute) CFT; see footnote \ref{relQFT}.

Consider the completely split scale-invariant SK geometry given by the isogenous fibration $(\til \pi, \til \cX)$ over $\cC$ whose generic fiber is $\til X_u$.
The results of section \ref{section 3} imply that $\til \cX$ is isotrivial.
The rigidity of isogenies mentioned above implies that the family of isogenies $f_u$ is constant over $\cC$, therefore, because $\til \cX$ is isotrivial, $\cX$ is as well.
Let us show this explicitly.

Being completely split and scale-invariant, the SK geometry on the isogenous fibration $\til\cX$  is isotrivial by the results of Section \ref{section 3}.
Thus, all its fibers $\til X_u$ are isomorphic to some fixed, completely split complex torus, $\til X$.
Our task is to show that the family of isogenies $f_u$ is actually constant on $\cC$; this will then imply that the ``physical" fibers $X_u$ over $\cC$ must be isotrivial, i.e., independent of $u \in \cC$.

Note that $\til X$ contains a \emph{completely split} basis in which the period matrix $\til \Pi$ takes the form, 
\begin{align}\label{Isplit_p-matrix}
    \til \Pi = \left ( \mathds 1_{r} \, \ \text{diag}(Z^1, \cdots, Z^r) \right ),
\end{align}
where the $Z^i \in \C$ with $\Im\, Z^i >0$ are the complex moduli of the rank-1 tori $\til X^i$ and are constant on $\cC$.
As described in appendix \ref{appA}, the isogeny $f_u$ implies $\til \Pi$ is related to the period matrix $\Pi_u$ of $X_u$ through
\begin{align}\label{isogeny_p-matrix}
    \til \Pi = G_u \Pi_u R_u^{-1}, \quad \qq{for all} \quad u \in \cC,
\end{align}
for some $G_u \in \GL(r,\C)$ and $R_u \in M(2r, \Z)$ with $|\mathrm{det}\, R_u| \in \N$.
Note that this relation is independent of the bases chosen for $\til X$ or $X_u$.
Since the matrix $R_u$ is integer valued, it cannot vary continuously with $u \in \cC$, so $R_u \doteq R$ is a constant, integer valued, invertible matrix (over $\Q$). 

Now consider the period matrices $\Pi_{u_1}$ and $\Pi_{u_2}$ defined at two points $u_1 \neq u_2$ in $\cC$ and with respect to a symplectic basis of each $(X_{u_i}, J)$.
Then each period matrix takes the form
\begin{align}
    \Pi_{u_i} = (\D \ \t(u_i))
\end{align}
where $\D=\text{diag}(d_1, \cdots, d_r)$ is the diagonal, integer matrix of \eqref{J Delta} and $\t(u_i)$ is the $r \times r$ symmetric matrix with positive definite imaginary part that represents the coupling of the low-energy $\U(1)^r$ gauge theory on $\cC$. 
By \eqref{isogeny_p-matrix}, each $\Pi_{u_i}$ is related to $\til \Pi$ according to
\begin{align}
    \til \Pi = G_{u_i}\Pi_{u_i} R^{-1},
\end{align}
and we can equate these two relations (since $\til \Pi$ is constant over $\cC$) to obtain 
\begin{align}
    G_{u_1} (\D \ \t(u_1)) R^{-1} = G_{u_2} (\D \ \t(u_2)) R^{-1}.
\end{align}
Since $R$ is invertible, this gives two constraints relating the $G_{u_i}$,
\begin{align}
    G_{u_1} \Delta = G_{u_2} \D \qand G_{u_1}\t(u_1) = G_{u_2} \t(u_2).
\end{align}
Focusing on the first relation, $\D$ is invertible since $\det \D = \prod_i d_i \neq 0$, so we find $G_{u_1} = G_{u_2} \deq G$ for some fixed $G \in \GL(r,\C)$, i.e., they are constant on $\cC$.
Since $G$ is invertible, the second relation implies the $\t(u_i)$ must be equal,
\begin{align}
    \t(u_1) = \t(u_2).
\end{align}
This proves the completely iso-split special K\"ahler geometry of $\cX$ on $\cC$ is isotrivial and the generic fiber is a fixed p.a.v.\  $(X,J)$ where $X$ is isogenous (as a complex torus) to a fixed, completely split complex torus, $\til X$.

\begin{example}\label{ex:d2} 
\ph{-}

\medskip
\noindent Recall that for genus-2 Riemann surfaces, the automorphism group is generically given by $\Z_2$ and is generated by the hyperelliptic involution. However, as we've seen in example \ref{ex:d6}, there exist surfaces $S$ for which the automorphism group enhances. The simplest case is given by $\mathrm{Aut}(S)=D_2\cong \Z_2\times\Z_2$. We will now examine this example in light of the discussion above.

Consider the family of smooth genus 2 Riemann surfaces with automorphism group $D_2\cong \Z_2\times \Z_2$. 
These can be described as
\begin{gather}
    X^{\mathbf{c}}[D_2] = \{ (x,y) \in \C^2: y^2 = (x^2-1) (x^4+c_2 x^2+c_0) \}.
\end{gather}
If $c_0=1$, the automorphism group enhances to $D_4$. 
By using considerations similar to the $X^c[D_6]$ case, we can show that the period matrix, in a canonical basis, takes the form
\begin{gather}
    (\D\ Z) = \begin{pmatrix}
        1 & 0 & x & y \\
        0 & 1 & y & x
    \end{pmatrix},
\end{gather}
with $x$ and $y$ functions of $c_0$ and $c_2$. 
This period matrix can be brought into split form 
\begin{gather}
    G \cdot (\D\ Z) \cdot R^{-1} =\begin{pmatrix}
        1 & 0 & \tfrac{1}{x-y} & 0   \\
        0 & 1 & 0 & \tfrac{1}{x+y} 
    \end{pmatrix},
\end{gather}
where
\begin{align}
    G &= \bpm
        \tfrac{1}{x-y} & -\tfrac{1}{x-y} \\
        \tfrac{1}{x+y} & \ph{-}\tfrac{1}{x+y}
    \epm,&
    R &= \bpm
        0 & \ph{-}0 & \ph{-}1 &-1 \\
        0 & \ph{-}0 & \ph{-}1 & \ph{-}1 \\
        1 &-1 & \ph{-}0 & \ph{-}0 \\
        1 & \ph{-}1 & \ph{-}0 & \ph{-}0
    \epm.
\end{align}
As before in example \ref{ex:d6}, this is not a symplectic basis change as $R\notin\mathrm{Sp}(4,\Z)$. However, this time we have $\det R=4$, meaning that this basis change does not induce an isomorphism of complex tori, but instead a degree $4$ isogeny of the form
\begin{gather}
     J(X^{\mathbf{c}}[D_2]) \sim E_{\t_1}\times E_{\t_2},
\end{gather}
where $\t_1=(x-y)^{-1}$ and $\t_2=(x+y)^{-1}$. 
\end{example}

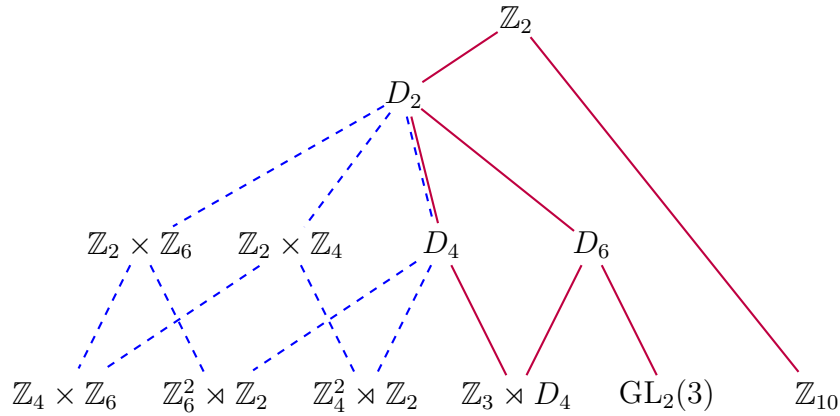
\begin{figure}
\centering
\begin{tikzpicture}
    \draw[thick,purple,shorten <=0.3cm,,shorten >=0.3cm] (0,0) -- (-1.5,-1);
    \draw[thick,purple,shorten <=0.3cm,,shorten >=0.4cm] (0,0) -- (4,-5);
    \draw[thick,blue,dashed,shorten <=0.3cm,,shorten >=0.3cm] (-1.5,-1) -- (-3,-3);
    \draw[thick,purple,shorten <=0.3cm,,shorten >=0.3cm] (-1.47,-1) -- (-0.97,-3);
    \draw[thick,blue,dashed,shorten <=0.3cm,,shorten >=0.3cm] (-1.53,-1) -- (-1.03,-3);
    \draw[thick,purple,shorten <=0.3cm,,shorten >=0.3cm] (-1.5,-1) -- (1,-3);
    \draw[thick,blue,dashed,shorten <=0.3cm,,shorten >=0.5cm] (-1.5,-1) -- (-5,-3);
    \draw[thick,blue,dashed,shorten <=0.3cm,,shorten >=0.4cm] (-5,-3) -- (-6,-5);
    \draw[thick,blue,dashed,shorten <=0.3cm,,shorten >=0.3cm] (-5,-3) -- (-4,-5);
    \draw[thick,dashed,blue,shorten <=0.5cm,,shorten >=0.6cm] (-3,-3) -- (-6,-5);
    \draw[thick,blue,dashed,shorten <=0.3cm,,shorten >=0.3cm] (-3,-3) -- (-2,-5);
    \draw[thick,blue,dashed,shorten <=0.3cm,,shorten >=0.3cm] (-1,-3) -- (-2,-5);
    \draw[thick,blue,dashed,shorten <=0.4cm,,shorten >=0.6cm] (-1,-3) -- (-4,-5);
    \draw[thick,purple,shorten <=0.3cm,,shorten >=0.3cm] (-1,-3) -- (0,-5);
    \draw[thick,purple,shorten <=0.3cm,,shorten >=0.3cm] (1,-3) -- (0,-5);
    \draw[thick,purple,shorten <=0.3cm,,shorten >=0.3cm] (1,-3) -- (2,-5);
    \node at (0,0){$\Z_2$};
    \node at (-1.5,-1){$D_2$};
    \node at (-5,-3){$\Z_2\times\Z_6$};
    \node at (-3,-3){$\Z_2\times\Z_4$};
    \node at (-1,-3){$D_4$};
    \node at (1,-3){$D_6$};
    \node at (-6,-5){$\Z_4\times\Z_6$};
    \node at (-4,-5){$\Z_6^2\rtimes \Z_2$};
    \node at (-2,-5){$\Z_4^2\rtimes \Z_2$};
    \node at (0,-5){$\Z_3\rtimes D_4$};
    \node at (2,-5){$\mathrm{GL}_2(3)$};
    \node at (4,-5){$\Z_{10}$};
\end{tikzpicture}
\caption{Hasse diagram showing the possible automorphisms of principally polarized abelian varieties of dimension two. Two groups are connected by a line if the upper group can be enhanced to the lower through specializing the parameters of a family of abelian varieties with the upper automorphism group. The line connecting them is \textcolor{purple}{solid purple} if this can be done for the Jacobian of a family of smooth hyperelliptic curves. The line is \textcolor{blue}{dashed blue} if this occurs for a product of elliptic curves instead. Note that $D_4 \cong \Z_2^2 \rtimes \Z_2$ and $D_2 \cong \Z_2 \times \Z_2$ can occur in both cases. 
}
\label{fig:aut_groups}
\end{figure}

This example has interesting consequences for the classification of rank-2 $\cN=2$ SCFTs. Of the possible genus-2 Riemann surfaces with an enhanced automorphism group (summarized in figure \ref{fig:aut_groups}), all of them, except for those with $\mathrm{Aut}(S)=\Z_{10}$, contain $\Z_2\times\Z_2$ as a subgroup. This immediately tells us that these cases are completely decomposable and, therefore, give rise to completely iso-split SK geometries which are isotrivial.
As such, these cases will be covered by the isotrivial classification presented in the upcoming work \cite{ACDMMTW24}. 
As for the $\Z_{10}$ model, we note that this still fits into the isotrivial regime, due to the fact that the only principally polarized abelian variety with automorphism group $\Z_{10}$ is the Jacobian of $y^2=x(x^5-1)$. We therefore conclude that all rank-2 theories with a generically enhanced automorphism group are isotrivial.

\acknowledgments

It is a pleasure to thank A. Bourget, S. Cecotti, M. Del Zotto, J. Grimminger, M. Lotito, M. Martone, and Y. Zekai for helpful conversations.  
We also thank the referee for insightful criticisms which were crucial for correcting our arguments.
This work is supported in part by DOE grant DE-SC1019775. 
RM is also supported by a Knut and Alice Wallenberg Foundation postdoctoral scholarship in mathematics.  
MW is supported by the National Research Foundation of Korea (NRF) Grants RS-2023-00208602 and NRF-2023R1A2C1004965.

\appendix

\section{Math appendix: complex tori and p.a.v.s}\label{appA}

In this appendix, we review the basics of complex tori and abelian varieties in order to elucidate some of the more subtle notions of ``split-ness'' that appear in the main body of the text.
We also anticipate that much of this information will be relevant for future work seeking to further classify physically allowed CB geometries, and in particular isotrivial geometries of arbitrary rank \cite{ACDMMTW24}.
For the benefit of the reader, table \ref{math_defs} provides a brief description of some terms we frequently use which characterize the ``splitting" properties of complex tori, p.a.v.s, and SK geometries.
Our presentation will mostly follow \cite{BL92,BL_tori,Lange23}. 

\begin{table}
\centering
\begin{tabularx}{\textwidth}{|c | X |} 
\hline\hline
 Term & Description \\ [0.0ex] 
 \hline\hline
 split complex torus & A complex torus which is isomorphic to the direct product of lower rank complex tori.  \\ \hline
 split SK geometry & An SK geometry where the generic fiber is a split complex torus. \\ \hline
 simple complex torus & A complex torus which contains no proper non-zero subtorus. \\ [0.0ex] \hline
 decomposable complex torus & A complex torus which is isogenous to the product of lower rank complex tori.  \\ \hline
 iso-split SK geometry & An SK geometry where the generic fiber is a decomposable complex torus. \\ \hline
 split (or reducible) p.a.v. & A p.a.v.\  which is isomorphic (as a p.a.v.) to the product of lower rank p.a.v.s. \\ \hline  \hline
\end{tabularx}
\caption{}
\label{math_defs}
\end{table}

\subsection{Complex tori}

A lattice $\L$ of rank $2g$ in a complex vector space $V \cong \C^g$ is the $\Z$-span of $2g$ complex vectors whose $\R$-span produces a basis for $V$ as a real vector space.
Given $\L$, we define the complex torus, $X=V/\Lambda$, which is compact and connected.
All compact connected complex Lie groups are complex tori;  as such we will use (complex) dimension and rank interchangeably for complex tori.
(In the body of the text we have called the rank $r$ instead of $g$.)

The period matrix of $X$ describes the embedding of $\L$ into $V$, and it can be defined as follows. 
Let $\{v_i:i=1,\ldots,g\}$ be a $\mathbb{C}$-linear basis for $V$ and $\{\s_a:a=1,\ldots,2g\}$ be a $\mathbb{Z}$-linear basis of $\Lambda$. 
We define the period matrix for $X$ to be the matrix $\Pi\in M(g\times 2g,\C)$ that encodes the embedding of the basis $\{\s_a\}$ of $\L$ in terms of the complex basis $\{v_i\}$ of $V$.
Concretely, we take $\Pi$ to satisfy the matrix relation, $\s^t = v^t \Pi$, which we write in terms of components as
\begin{align}\label{P_matrix}
    \s_a = v_j \Pi^j_{\ a}, \qquad \Pi = \begin{pmatrix}
        \Pi_1^1 & \cdots & \Pi_{2g}^1 \\
        \vdots &  \ddots & \vdots \\
        \Pi_1^g & \cdots & \Pi_{2g}^g
    \end{pmatrix} \deq (\sfA \ \sfB).
\end{align}
where $\sfA$ and $\sfB$ are both $g\times g$ matrices.
That $\L$ is of rank $2g$ means the $2g \times 2g$ matrix $(\Pi^t\ \bar \Pi{^t})$ is nonsingular, i.e., $\det \bigl(\Pi^t\ \bar \Pi{^t} \bigr) \neq 0$.
Conversely, any matrix $\Pi\in M(g\times 2g,\C)$ satisfying this condition describes a complex torus, \cite[prop.\ 1.1.5]{Lange23}.  
Thus, by construction, the complex torus $X$ is fully specified by the period matrix and can be described by $\C^g/\Pi \Z^{2g}$. 

As the name suggests, we can also describe the period matrix in terms of the periods, $\int_{\s_a} \w^j$, formed between basis elements $\{\w^i \}$ of the $(1,0)$-cohomology and $\{\s_a \}$ of the $1$-homology of $X$.
A basic result (\cite[lemma 1.1.3]{Lange23}) identifies the lattice $\L$ with the $1$-homology lattice of $X$.
For a basis $\{ \s_a \}$ of $H_1(X,\Z)\cong_{\Z}\L$, let $\{ x^a \}$ be the corresponding coordinates on $V$ and define the dual $\R$-basis $\{\diff x^a\}$ of $V^*$ which form a $\Z$-basis of $1$-forms on $H^1(X,\Z)$ satisfying
\begin{align}\label{real_periods}
    \int_{\s_a} \diff x^b = \d^b_a .
\end{align}
For a generic $\C$-basis $\{v_i\}$ of $V$, the corresponding dual $\C$-basis $\{\w^i \deq \diff v^i\}$ of $V^* \cong_{\C} H^{(1,0)}(X)$ constitute a basis for the global $(1,0)$-forms on $X$, and \ref{P_matrix} implies they are related to $\{\diff x^a\}$ via the period matrix,
\begin{align}\label{dual C-basis}
    \w^i = \Pi^i_{\ a} \diff x^a.
\end{align}
From \eqref{real_periods} it follows that 
\begin{align}\label{period_mat_2}
    \Pi^j_{\ a} &= \int_{\s_a}\w^j & 
    \qand & & 
    \sfA_i^j &= \int_{\a_i} \w^j , &
    \sfB^{ij} &= \int_{\b^i} \w^j .
\end{align}
where $\{ \a_i \deq \s_i, \ \b^i \deq \s_{i+g} \}$.

Let $X = V/\L$ and $X' = V'/\L'$ be complex tori of complex dimensions $g$ and $g'$, respectively.
A homomorphism $f: X \to X'$ is a holomorphic map preserving the group structure of $X$, $X'$ as complex abelian groups.%
\footnote{Generic holomorphic maps $h:X \to X'$ can be decomposed as $h(x)=f(x)+h(0)$, where $f: X \to X'$ is a unique homomorphism.}
Since $V$ and $V'$ are the universal coverings of $X$ and $X'$, respectively, covering theory indicates that every such homomorphism lifts to a unique $\C$-linear map $F:V\to V'$ such that $F(\L) \subseteq \L'$.
This gives an injective homomorphism of abelian groups $\mathsf{Hom}(X, X') \to \mathsf{Hom}_{\C}(V, V')$ that associates $f \mapsto F$.
Furthermore, since $F(\L) \subseteq \L'$, we also get an injective homomorphism $\mathsf{Hom}(X, X') \to \mathsf{Hom}_{\Z}(\L, \L')$ mapping $f \mapsto F_{\L} \deq F|_{\L}: \L \to \L'$.
So knowledge of the lattice map $F_\L$ is sufficient to reconstruct the torus homomorphism $f$.

Now let $\Pi \in M(g \times 2g, \C)$ and $\Pi' \in M(g' \times 2g',\C)$ be the period matrices of $X$ and $X'$ written with respect to some bases for $V$, $\L$ and $V'$, $\L'$, and consider a homomorphism $f: X \to X'$.
With respect to these bases, the unique (analytic) map $F\in \mathsf{Hom}_{\C}(V,V')$ associated to $f$ and its restriction $F_{\L} \in \mathsf{Hom}_{\Z}(\L,\L')$ are represented by matrices $G \in M(g' \times g,\C)$ and $R \in M(2g' \times 2g, \Z)$, respectively.%
\footnote{If $\{ \s_a \}$ is a basis for $\L$ and $\{ \s'_{a'} \}$ the  basis for $\L'$, then $R$ is defined to be the matrix representation of the $\Z$-linear action of $F_{\L}$ on the row vector $\s^t$: $F_{\L}(\s_a) \deq \s'_{b'} R^{b'}_{\ a}$.
Likewise, if $\{ v_i \}$ is a basis for $V$ and $\{ v'_{i'} \}$ is a basis for $V'$, then $G$ is the matrix representation of the $\C$-linear action of $F$ on $v^t$: $F(v_i) \deq v'_{j'} G^{j'}_{\ i} $.
This implies the corresponding matrices implementing basis changes of $\L$ and $V$ that map $\Pi \mapsto \Pi'$ are given by $R^{-t}$ and $G^{t}$, respectively. }
The condition $F(\L) \subseteq \L'$ then implies $F(m) = F_{\L}(m) \in \L'$ for all $m \in \L$ from which one obtains the matrix relation,
\begin{align}\label{CT_homo}
G \Pi = \Pi' R.	
\end{align}
Conversely, any two matrices $G \in M(g' \times g,\C)$ and $R \in M(2g' \times 2g, \Z)$ satisfying \eqref{CT_homo} define a homomorphism $X \to X'$.

When $f$ is an isomorphism, $g = g'$, $G \in \GL(g,\C)$, and $R \in \GL(2g, \Z)$, so the period matrices satisfy
\begin{equation}\label{PM_iso}
\Pi' = G \Pi R^{-1}. 
\end{equation}
Such transformations of $X = V/\L$ describe basis changes of $V$ and $\L$, reflecting the fact that our choice of bases to define $\Pi$ in \eqref{P_matrix}, and therefore $X$ itself, was not unique.
Due to the relation \eqref{period_mat_2}, we will often describe basis changes of $\Pi$ as resulting from a basis changes in the $(1,0)$-cohomology and $1$-homology of $X$.
It is useful to note that there always exists a basis such that $\Pi=(\mathds 1_{g}, Z)$ where $\det \Im Z \neq 0$.

Letting $\Pi = (\sfA\ \sfB)$ and $\Pi' = (\sfA'\ \sfB')$ as in \eqref{period_mat_2}, one can deduce the transformation of the quantity $\sfB^t \sfA^{-t}$ under the homology lattice basis change $R^{-t}$.
Define $R^{-t} \doteq \bspm d & c \\ b & a \espm \in \GL(2g,\Z)$ in $g \times g$ block form, then
\begin{align} \label{PM_basis_trans} 
\sfB^t \sfA^{-t} \mapsto (\sfB')^t (\sfA')^{-t} = \left ( a \sfB^t \sfA^{-t} + b \right )\left ( c \sfB^t \sfA^{-t} + d \right )^{-1} ,
\end{align}
from which \eqref{t to tt} follows.

\subsubsection{Split complex tori}

A complex torus $X$ of rank $g$ is said to be \emph{split} if it is isomorphic to the direct product of $r$ lower-rank complex tori $X_i$, 
\begin{align}
    X \cong_{\C} X_1 \times \cdots \times X_r,
\end{align} 
where each torus $X_i$ for $i = 1, \ldots, r$ has rank $g_i$ which satisfy $g = g_1 + \cdots + g_r$.
If $X$ splits, there necessarily exists a basis of the fiber 1-homology and the fiber $(1,0)$-cohomology such that period matrix can be written in the block diagonal form,
\begin{align}\label{split_ct_1}
    \Pi = \bpm \Pi_1 & &  \\
    & \ddots & & \\
    & & \Pi_r \epm ,
\end{align}
where each $\Pi_i$ is a $g_i \times 2 g_i$ complex matrix satisfying $\det \bigl(\Pi^t_i\ \bar \Pi{^t_i} \bigr) \neq 0$.
Note that we only demand that the splitting isomorphism respect the complex manifold and abelian group structure of the torus.
Furthermore, our definition of split allows two torus factors $X_i, X_j$ for $i \neq j$ to be isomorphic and we assume each torus factor $X_i$ is \emph{non-split}, i.e., that it does not itself split into lower-rank sub-tori.

From the fact that the period matrix of each non-split torus $X_i$ can be put into the form $\Pi_i = (\mathds 1_{g_i}\ Z^i)$ where $Z^i \in M(g_i,\C)$ such that $\det \Im Z^i \neq 0$, we can refine the basis choices of $H_1(X,\Z)$ and $H^{(1,0)}(X)$ such that $\Pi$ takes the form
\begin{align}\label{split_ct_2}
    \Pi &= (\mathds 1_g\  Z), & Z &= \bpm Z^1 & & \\
    & \ddots & & \\
    & & Z^r \epm .
\end{align}
We will call such a basis a \emph{split basis for} $X$. 

When $X$ splits with $g_i=1$ for all $i$, then the complex torus is isomorphic to the product of $r=g$ rank-$1$ complex tori, i.e., elliptic curves, and we say it is \emph{completely split}. 
Relative to \eqref{split_ct_2}, in the completely split case we can always further refine the split basis of the period matrix for $X$ so that $Z^i$ satisfies $\Im \, Z^i>0$ and $(1, Z^i)$ spans the rank-$2$ sublattice $\L_i \subset \L$ defining $X_i$ as the rank-$1$ complex torus $\C/\L_i$.
This form of the period matrix corresponds to choosing a basis $\{ \w^i \}$ of the $(1,0)$-cohomology and a canonical basis $\{\a_i, \b^i \}$ of $1$-homology cycles of $X$ such that it decomposes into a direct sum of the corresponding $(1,0)$-cohomology and $1$-homology groups of each $X_i$,
\begin{align}\label{CT_csplit_basis}
   \sfA &= \int_{\a_i} \w^j = \d^j_i,& 
   \sfB &= \int_{\b^i} \w^j = \d^{ij} Z^i.
\end{align}

\subsubsection{Isogenies and decomposable complex tori}\label{AppA:isog_ct}

An {\it isogeny} between two complex tori $X$ and $X'$ is a surjective homomorphism $f:X\to X'$ with finite kernel. 
This means that $X$ and $X'$ have the same rank and $X$ is a ${\rm deg}\, f \doteq |\ker f|$ cover of $X'$.
While isogenies are not isomorphisms unless ${\rm deg} f=1$, they are ``almost'' isomorphisms in the following sense. 
Define the exponent of the isogeny $f$ to be the smallest positive integer $n$ such that $n x =0$ for all $x\in\ker f$, i.e., the smallest $n$ such that as a group $\ker f \subset (\Z_n)^{2g}$ where $g$ is the rank of $X$.
Then, given an isogeny $f:X\to X'$ with exponent $n$, there exists a unique \emph{dual isogeny} $\hat f:X'\to X$ (up to isomorphism) such that $\hat f\circ f= n\, \mathrm{id}_X$ and $f\circ \hat f = n\, \mathrm{id}_{X'}$. 
(Thus ${\rm deg}\hat f \cdot {\rm deg}f = n^{2g}$.)
It follows that isogenies define an equivalence relation on the set of complex tori. 
Tori that belong to the same equivalence class are said to be {\it isogenous}, denoted by $X\sim X'$.

It will be useful for practical purposes to understand isogenies at the level of the $1$-homology lattice and period matrices. 
Recall that a homomorphism between complex tori $f:V/\Lambda\to V'/\Lambda'$ is induced by another homomorphism $F:V\to V'$ satisfying $F(\L)\subseteq \L'$. 
Restricting to the case where $f$ is an isogeny, it follows that $\dim_{\C} X = \dim_{\C} X' = g$ and $F(\L)$ is a maximal-rank (i.e., finite index) sublattice of $\L'$.
Indeed, these conditions define an isogeny, and the index $|\L'/F(\L)| = {\rm deg} f$.
In terms of period matrices $\Pi, \Pi'\in M(g\times 2g,\C)$ for $V/\Lambda$ and $V'/\Lambda'$ respectively, $f$ being an isogeny translates to the existence of matrices $G\in \GL(g,\C)$ and $R\in M(2g,\Z)$ such that
\begin{align}\label{isog_pmatrix}
   G \Pi &= \Pi' R, &
   |\!\det R| & = {\rm deg} f.
\end{align}

A complex torus is \emph{decomposable} if it is isogenous to a product of lower rank complex tori.
The decomposition theorem for complex tori \cite[theorem 1.7.5]{BL_tori} states that every complex torus has a unique (up to isogeny) decomposition into indecomposable tori: given an arbitrary complex torus $X$, there exists an isogeny
\begin{gather}\label{eq:decomp}
    X\sim X_1^{n_1}\times \cdots \times X_r^{n_r},
\end{gather}
where the $n_i$ are positive integers and the $X_i$ are pairwise non-isogenous indecomposable complex tori which are uniquely determined up to isogenies and permutations.

(SK geometries whose generic fiber is decomposable are what we have called ``iso-split'' in section \ref{section 4}.
When the generic fiber decomposes as in \eqref{eq:decomp} with $\sum_i n_i=\dim_\C X$, we say the SK geometry is completely iso-split, since then each of the $X_i$ factors have dimension 1.)

Note that by going to an isogenous cover if necessary, any particular factor on the right side of \eqref{eq:decomp} appears as a subtorus of $X$.
A complex torus is \emph{simple} if it has no proper non-zero complex subtorus. 
Clearly, if a complex torus is simple then it must be indecomposable, but the converse need not be true.
The following provides an example of a family of rank-2 complex tori that are indecomposable, but not simple.

\begin{example}\label{shafarevich}

{\bf Rank-2 tori which are indecomposable, but not simple.}%
\footnote{This example is adapted from \cite[example 8.3, p.\ 158]{Shaf13}.}

\medskip
\noindent Consider the rank-2 torus $X = V/\Pi \Z^4$ with $V=\C^2$ and period matrix
\begin{align}
    \Pi = \bpm 1 & 0 & i & \a \\ 0 & 1 & 0 & i \epm ,
\end{align}
where $\a$ is an arbitrary complex number.
Define $X_1 \doteq V_1/\Pi_1 \Z^2$ with $V_1 = \C$ and $\Pi_1 = (1\ i)$.   
Take $V_1 \subset V = \C^2$ to be the complex subspace consisting of the first factor which lies in the real span of the lattice basis elements $\bspm 1\\ 0\espm$ and $\bspm i\\ 0\espm$.
Then $X_1$ is a proper subtorus of $X$, so $X$ is not simple.
If $X$ were decomposable, then there would exist a second subtorus, $X_2$, such that $X$ is isogenous to $X_1\times X_2$.  
Any such second complex subtorus can be written $X_2 = V_2/\Pi_2 \Z^2$ with $\Pi_2 = (1\ \l)$ for some $\Im\l\neq0$ and $V_1 \neq V_2 \subset V$.
A complex basis for $V_2 \subset V$ must be a vector $e \in \Pi \Z^4$ 
\begin{align}
    e = \Pi \bspm a\\ b\\ c\\ d \espm
    = \bpm a+ic+\a d\\ b+ id \epm
\end{align}
where $a,b,c,d$ are integers such that $e\notin V_1$, i.e., such that $b+id \neq 0$.
Now, for $X_2$ to be a complex subtorus of $X$ we also need that $\l e \in \Pi \Z^4$, or
\begin{align}
    \bpm \l a + i \l c + \a \l d \\
    \l b + i \l d \epm =
    \bpm a' + i c' + \a d' \\
    b' + i d' \epm
\end{align}
for some integers $a',b',c',d'$.
The second component of this equation requires that $(b+id) \l \in \Z + i \Z$ which implies that $\l \in \Q(i)$ (since $b+id\neq 0$).
Also, since $\Im \l \neq0$, it implies that not both $d$ and $d'$ can vanish.
The first component then requires $\a\in \Q(\l,i)=\Q(i)$.
Therefore, if either the real or imaginary part of $\a$ is irrational, then no subtorus $X_2$ exists, and $X$ is indecomposable.
Conversely, if $\a \in \Q(i)$, say $\a = (p/q) + i (r/s)$ for some integers $p,q,r,s$ ($q$ and $s$ non-zero), then $X$ is decomposable.
For instance, $\l= -i q/s$ is a solution for $X_2$ (with $a=b=0$, $c=-r$, $d=s$, $a'=0$, $b'=q$, $c'=-p$, $d'=0$).

\end{example}

\subsection{Polarized abelian varieties (p.a.v.s)}\label{p.a.v.s}

Complex tori that can be embedded in projective space are called \emph{abelian varieties}.
By the Kodaira embedding theorem, only those complex tori that admit a positive holomorphic line bundle are algebraic in this way.
All rank-$1$ tori are abelian varieties as they can be described by elliptic curves in projective space, but starting at rank $2$ there exist complex tori which are not abelian varieties.

Analytic equivalence classes of holomorphic line bundles on a complex torus are given by their Chern classes, $c_1(L)$, which are $(1,1)$ cohomology classes that are also integral, i.e., an element of $H^{(1,1)}(X) \cap H^2(X,\Z)$.
On a rank-$g$ complex torus $X=V/\L$ a representative of this cohomology class can be taken to be constant by translation invariance.
Thus the coefficients of $c_1(L)$ relative to a complex basis of $V$ define a hermitian form $H:V\times V \to \C$ such that $\Im H(\L,\L) \in \Z$.
Define the real skew-symmetric bilinear form on $V$ by $J \doteq \Im H$.   
Then hermiticity of $H$ implies $J(iv,iw) = J(v,w)$ and $H(v,w) = J(iv,w) + iJ(v,w)$.
In a real basis of $V$ dual to a basis of $\L$, as in \eqref{real_periods}, the coefficients of $J$ are a non-degenerate, integer, skew-symmetric $2g\times 2g$ matrix, $J$, called a \emph{symplectic pairing} on the lattice $\L$.
Note that such $J$'s are in 1-to-1 correspondence with \emph{non-degenerate} (but not necessarily positive) $H$'s with integral imaginary part.

A complex torus $X=V/\L$ is an abelian variety if it admits a \emph{positive} line bundle, which translates into the existence of a positive-definite hermitian form $H$ on $V$.
By virtue of $H$ being positive-definite, $J$ satisfies certain compatibility conditions with the period matrix of $X$.
These conditions are the so-called Riemann conditions, or Riemann bilinear relations \cite[sec.\ 2.1.5]{Lange23}. 
$J$ is then called a \emph{polarization} on the abelian variety $X$, and we call the pair $(X,J)$ a \emph{polarized abelian variety}, or \emph{p.a.v.}\ for short.
Before describing the form of the Riemann conditions that are most useful for our purposes, we make a brief digression to describe some basic properties of symplectic pairings.

\subsubsection{Symplectic lattices}

A symplectic pairing $J$ on a lattice $\L$ is a non-degenerate, integral, skew-symmetric bilinear form $J: \L \times \L \to \Z$.
When a lattice $\L$ is equipped with such a pairing, we call the resulting pair $(\L, J)$ a \emph{symplectic lattice}.
Relative to some basis $\{\s_a\}$ of $\L$, let $J_{ab} \deq J(\s_a, \s_b)$ be the matrix representation of $J$.
Then, $J(u,w) = u^a J_{ab} w^b$ for all $u,w \in \L$.
(Note that we are abusing notation by letting $J$ represent both the matrix form of the symplectic pairing $J$ written with respect to some basis of $\L$ and the pairing $J$ itself.)
Under arbitrary basis transformations $M: \s \to \s'\deq M\s$, $\s_a \mapsto \s'_a \deq M_a^{\ b} \s_b$, defined by $M \in \GL(2g,\Z)$, the matrix elements of $J$ transform according to $J_{ab} \mapsto  J'_{ab} = (M\cdot J \cdot M^t)_{ab}$.

There exists a basis $\{\a_i,\b^i\}$ of any symplectic lattice $\L$ such that $J$ can be written as a $2g \times 2g$ skew-diagonal matrix as in \eqref{J Delta} with $\D$ given in \eqref{tau Delta} \cite[lemma, ch.\ 2.6]{GH78}.
The positive integer entries $d_i$ satisfying $d_i | d_{i+1}$ are the \emph{invariant factors} of $J$ and they uniquely characterize it up to changes of lattice basis.
It follows that $\det J = (\text{Pf}J)^2 = (\prod_i d_i)^2$.
When all $d_i =1$, the pairing is said to be \emph{principal}.
We refer to a basis $\{ \a_i, \b^i \}$ for which the pairing takes the form \eqref{J Delta} as a \emph{symplectic basis} of $J$.
The $\{\a_i\}$ and $\{\b^i\}$ elements generate lagrangian sublattices with respect to $J$.

Two symplectic lattices are isomorphic if there exist bases of each lattice in which the matrix form of their symplectic pairings numerically agree.
Based on the uniqueness of the Smith normal form for each pairing, this is true iff both pairings have the same invariant factors.
Automorphisms $M \in \mathsf{Aut}(\L,J) \subset \GL(2g,\Z)$ of a symplectic lattice, $(\L,J)$, are those transformations that preserve the numerical form of $J$ with respect to a given basis: $M^t J M = J$.
This is the definition of the matrix group $\Sp_J(2g,\Z)$; when $J$ is principally polarized and in standard form, $J = \mathds 1_g \otimes \e$, this is the usual integral symplectic group, $\Sp(2g,\Z)$.

\subsubsection{Riemann conditions}

Let $X=V/\L$ be a rank-$g$ complex torus with period matrix $\Pi$ and consider a symplectic pairing $J$ defined on $\L$.
The Riemann conditions state that $X$ is an abelian variety if and only if the following conditions are satisfied,
\begin{align} \label{RCII} 
\Pi \cdot J^{-1} \cdot \Pi^t &= 0, &  i \Pi \cdot J^{-1}\cdot \bar \Pi^t &>0, &
\end{align}
where $\bar \Pi$ is the complex conjugate of $\Pi$ and the second condition denotes the requirement that the real matrix $i \Pi \cdot J^{-1}\cdot \bar \Pi^t$ be positive definite.
These follow straightforwardly from the conditions that $J(iv,iw)=J(v,w)$, that $J=\Im\, H$, and that $H$ is positive-definite.

Although these conditions are basis independent, we can specialize to a particular basis of $X$ defined relative to a choice of symplectic basis of the pairing $J$ to put them in a simple form.
In particular, given a choice of $J$ symplectic basis of $\L$ as in \eqref{J Delta}, there exists a basis $\{ \w^i\}$ of the $(1,0)$-cohomology of $X$ such that the period matrix of $X$ takes the form
\begin{align}\label{PM_Delta_tau}
    \Pi &= (\sfA\ \sfB) &
    & \text{with} &
    \sfA_i^j &=\int_{\a_i} \w^j =\D^j_i, &
    \sfB^{ij} &= \int_{\b^i} \w^j \doteq \t^{ij},
\end{align}
where $\t^{ij}$ is a $g \times g$ complex matrix and $\D$ is the integer diagonal matrix of invariant factors given by the polarization as described in \eqref{J Delta}, \eqref{tau Delta}.
The conditions in \eqref{RCII} written in these bases translate to the requirements
\begin{align} \label{RCIII} 
\t &= \t^t, & \Im \, \t >0, 
\end{align}
(which are the defining conditions for $\t$ to be valued in the rank-$g$ Siegel upper-half space) for $X$ to be an abelian variety.
When $1$-homology and $(1,0)$-cohomology bases of a p.a.v.\ $(X,J)$ are chosen such that the period matrix satisfies \eqref{PM_Delta_tau}, \eqref{RCIII} with $J = \e \otimes \D$, we call the pair of basis choices a \emph{symplectic basis of} $(X,J)$.

We will often refer to $\t$ as the \emph{(complex) modulus} of the p.a.v.\ $(X,J)$, or simply of $X$ when the polarization is understood.
Note that $\t$ is not uniquely determined by going to a symplectic basis. 
This is because a basis change of the form \eqref{PM_iso} with $R^{-t} \doteq \bspm d & c \\ b & a \espm \in \Sp_J(2g,\Z)$ preserves the polarization and takes the modulus to
\begin{align} \label{pav_sb_trans} 
\t \mapsto \t' 
= (a\t + b\D) (c\t + d\D)^{-1} \D ,
\end{align}
by \eqref{PM_basis_trans}.


  
\subsubsection{Isogenies and isomorphisms of p.a.v.s} \label{maps_of_pavs}

Consider two p.a.v.s $(X, J)$ and $(X',J')$ with polarizations $J$ and $J'$ and lattices $\L$ and $\L'$, respectively.
A \emph{homomorphism of p.a.v.s} $f: (X,J) \rightarrow (X',J')$ is a homomorphism of their complex tori $f: X \rightarrow X'$ such that the polarizations are preserved, i.e., $J$ is the pullback of $J'$.
We say that $J$ is the polarization \emph{induced by} $J'$ \emph{via} $f$.
In particular, $f$ is a homomorphism of p.a.v.s when the period matrices $\Pi$ and $\Pi'$ of $X$ and $X'$ are related as in \eqref{CT_homo} and $ J = F_\L^*J' = J' \circ F_\L$ for the unique mapping $F_\L \in \mathsf{Hom}_\Z(\L,\L')$ associated to $f$.
Explicitly, let $G \in M(g'\times g, \C)$ be the matrix representation of $F\in \mathsf{Hom}_{\C}(V, V')$ and $R\in M(2g' \times 2g,\Z)$ be the matrix representation of $F_{\L} = F|_{\L}$ so the period matrices satisfy $G \Pi = \Pi' R$.
Then, the induced polarization is $J = R^t \cdot J' \cdot R$.
Since $J$ is nondegenerate, the rank of $R$ must be maximal, so $\ker f$ must be finite.
In particular, if the dimensions of two p.a.v.s are the same, any homomorphism between them must be an isogeny of their underlying tori, in which case $R$ and $G$ are invertible.

Conversely, given a homomorphism $f: X \to X'$ of complex tori with finite kernel and $J'$ a polarization on $X'$, the induced symplectic pairing $F_\L^* J'$ on $\L$ defines a polarization on $X$ (because the pullback of the associated hermitian form remains positive-definite) rendering it an abelian variety. 
Therefore, a complex subtorus of an abelian variety is an abelian variety.
Also, a complex torus isogenous to an abelian variety is an abelian variety.

When a homomorphism $f: (X,J) \rightarrow (X',J')$ of p.a.v.s is also surjective, so $f: X \to X'$ is an isogeny of complex tori, then we say $(X,J)$ and $(X',J')$ are \emph{isogenous as p.a.v.s}.
For isogenies with ${\rm deg} \, f >1$, the invariant factors of the polarizations are never preserved.
To see this, let $R \in M(2g,\Z)$ be the matrix defining the isogeny in \eqref{isog_pmatrix} so the relation $J' = R^{-t} \cdot J \cdot R^{-1}$ holds for an isogeny of p.a.v.s.
For ${\rm deg} \, f = |\!\det R| >1$, it follows that $\det J' < \det J$ which means $J$ and $J'$ don't have the same invariant factors, and the product of the invariant factors always decreases under a non-trivial isogeny of p.a.v.s.
This result is consistent with the fact that all polarizations on a p.a.v.\ are induced by principal polarizations via isogenies \cite[prop 2.1.2]{Lange23}.

A morphism of p.a.v.s $f:(X,J) \to (X',J')$ is an \emph{isomorphism} when $G \in \GL(g,\C)$ and $R \in \GL(2g,\Z)$.
Uniqueness of the Smith normal form for the polarizations together with $J = R J' R^t$ implies $J$ and $J'$ have the same invariant factors, i.e., $\D = \D'$.
Thus, when two p.a.v.s are isomorphic there are $1$-homology bases in which $J=J'$ as matrices.
Relative to such bases, the residual isomorphisms $M \in \GL(2r,\Z)$ preserving the matrix form of $J$, $M^t J M = J$, are the automorphisms of the symplectic lattice, $M \in \mathsf{Aut}(\L,J) \cong \Sp_J(2g,\Z)$.
When $(X',J') = (X,J)$, so $G \in \GL(r,\C)$ and $R^{-t} \in \Aut(\L,J)$  preserve the period matrix and polarization of $(X,J)$, such isomorphisms define the automorphism group, $\mathsf{Aut}(X,J)$, of $(X,J)$.
Unlike the case of complex tori which can have automorphisms of infinite order, this is a finite group \cite[corollary 5.1.9]{BL92}.
For example, the possible automorphism groups of rank-2 principally polarized abelian varieties are shown in figure \ref{fig:aut_groups}.

It might be clarifying to note that it is possible for an abelian variety to admit inequivalent polarizations with the same invariant factors.%
\footnote{Example \ref{ex:d6} provides an instance of this.}
Consider the two p.a.v.s $(X, J_1)$ and $(X, J_2)$ for an abelian variety $X$ where each $J_i$ has the same invariant factors as a symplectic pairing.
Upon going to a symplectic basis of each p.a.v., the polarizations $J_i$ will have the same matrix form, $J=\e \otimes \D$, but the period matrices take the form $\Pi_1 = (\D, \t_1)$ and $\Pi_2 = (\D, \t_2)$ where the $\t_i$ may be in different $\Sp_J(2g,\Z)$ orbits under the action \eqref{pav_sb_trans}.
Another way of saying this is to note that because the period matrices $\Pi_i$ describe the same underlying complex torus $X$, there necessarily exists $R \in \GL(2g,\Z)$ and $G \in \GL(g,\C)$ such that $\Pi_2 = G \Pi_1 R^{-1}$, but there is no reason that $J_2 = R^{-t} \cdot J_1 \cdot R^{-1}$ need be true, i.e., that the basis changes of $X\cong_{\C} V/\L$ must also identify the different polarizations it can be equipped with.

\subsubsection{Decomposable, simple abelian varieties and split p.a.v.s}

The Poincar\'e complete reducibility theorem for abelian varieties \cite[theorem 2.4.25]{Lange23} states that every abelian variety has a unique (up to isogeny) decomposition into \emph{simple} abelian varieties.
This is a stronger version of the decomposition theorem for complex tori \eqref{eq:decomp} where the tori in the product decomposition were only guaranteed to be \emph{indecomposable}.
In particular, it implies that simplicity and indecomposability are equivalent for abelian varieties.

As pointed out in the previous section, an isogeny of p.a.v.s with non-trivial kernel doesn't preserve the invariant factors of a polarization, and so can never be an isomorphism of p.a.v.s.
Since isomorphisms of p.a.v.s are required to preserve the invariant factors of the polarizations involved, they are a stronger equivalence relative to isogenies of p.a.v.s.
Isomorphisms of p.a.v.s are thus a stronger equivalence relative to isogenies of p.a.v.s because they are required to preserve the invariant factors of the polarizations involved.
In particular, we say that a p.a.v.\ $(X,J)$ \emph{splits} (or is \emph{reducible}) when it is isomorphic (as a p.a.v.) to a direct product of $r$ lower-rank p.a.v.s,
\begin{align}\label{split_pav}
    (X,J) = (X_1, J_1) \times \cdots \times (X_r, J_r)
\end{align}
with $J = p_1^*J_1 \oplus \cdots \oplus p_r^*J_r$, where $p_i : X \to X_i$ is the projection onto the $i$th factor.
Going to symplectic bases of each $(X_i,J_i)$ factor as in \eqref{PM_Delta_tau} implies the existence of a basis in which the period matrix of $(X,J)$ takes the split form $\Pi = (\til\D, \t)$ with $\t = \t_1 \oplus \cdots \oplus \t_r$ and $\til\D = \D_1 \oplus \cdots \oplus \D_r$.
Note, however, that this is not necessarily a symplectic basis for $(X,J)$ since $\til\D$ need not be in invariant factor form.  
Even though it is diagonal, $\til \D = {\rm diag} ( \til d_1, \ldots, \til d_g )$, with positive integer diagonal entries $\til d_i$, they need not satisfy the divisibility requirements $\til d_i | \til d_{i+1}$ required of invariant factors.
A further basis change to Smith normal form is required; see, for example, \cite[appendix B]{Argyres:2022kon}.
Note that this further basis change need not preserve the split form of $\Pi$; that is, the split basis need not be simultaneously a split and symplectic basis for $(X,J)$ to split as a p.a.v.

To be clear, a p.a.v.\ $(X,J)$ where $X$ splits as a complex torus need not split as the p.a.v.\ $(X,J)$.
Indeed, examples \ref{ex:d6} and \ref{ex:d2} provide instances of this.
But, given any p.a.v.\ $(X,J)$ where $X$ splits as a complex torus, it is always possible to construct a corresponding split p.a.v., $(X,J')$.
In particular, if $X=X_1 \times \cdots \times X_r$, define $J' \doteq J_1 \oplus \cdots \oplus J_r$ where $J_i \deq J|_{X_i}$ are polarizations obtained by restricting $J$ to each torus sub-factor, i.e., via the polarization induced from the embeddings $f_i: X_i \hookrightarrow X$.%
\footnote{That these are polarizations follows from the fact that hermitian form $H_J$ corresponding to $J$ remains positive-definite after being restricted to each torus sub-factor, \cite[prop. 2.1.1]{Lange23}.}
Then $(X,J')$ is a split p.a.v., though it need not be isomorphic to $(X,J)$. 

\subsubsection{Some results on completely split abelian varieties}
\label{app Jac}

If $(X,J)$ is a p.a.v.\ in which $X$ \emph{completely} splits, then $(X,J')$, defined above, is a completely split p.a.v.; furthermore, in this case one can take $J'$ to be a principal polarization, since rescaling the $J_i$ factors by positive integers does not affect the positivity of the hermitian form corresponding to $J$.
In other words, given a completely split p.a.v., there is always an isogeny to a completely split p.a.v. in which the image polarization is principal.
This suggests that the class of completely split abelian varieties is more amenable to a general classification and construction than are the completely split complex tori.
This section collects a few useful results illustrating this.

There is a particular subclass of completely split abelian varieties for which the distinction between isogeny and isomorphism becomes {\it almost} insignificant.

\begin{theorem}
    {\bf \cite[corollary 10.6.3 and exercise 10.5]{BL92}.}  
    If $X$ is an abelian variety which completely iso-splits to the product $\times_{i=1}^g E$ with a fixed elliptic curve $E$ which has complex multiplication, then $X$ completely splits as $X = \times_{i=1}^g E_i$ with $E_i$ elliptic curves isogenous to $E$ \cite{Shioda74, Schoen92}. 
\end{theorem}

\noindent 
\emph{Complex multiplication} is defined to mean that the endomorphism ring of the elliptic curve, ${\rm End}(E)$, has rank greater than 1 as a $\Z$-module. 
We recall here the basic properties of the endomorphism rings of complex elliptic curves.
The endomorphism algebra, ${\rm End}_\Q(E) \doteq {\rm End}(E) \otimes \Q$ of an elliptic curve is an imaginary quadratic field $\Q(\sqrt{-m})$ where $m$ is a square-free non-negative integer.
This happens whenever the modulus $\t\in \Q(\sqrt{-m})$ for some integer $m>0$; if there is no such $m$ then ${\rm End}_\Q(E) = \Q$.
Thus ${\rm End}_\Q(E)$ has dimension 2 when $m>0$.
The endomorphism ring, ${\rm End}(E)$ is then an \emph{order} $\mathfrak{o} \subset {\rm End}_\Q(E)$, that is, a subring which is a module over $\Z$ of maximal rank (i.e., 2 when $m>0$ and 1 when $m=0$).
Thus $E$ has complex multiplication iff its endomorphism algebra is $\Q(\sqrt{-m})$ with $m>0$.

In rank 2, Ruppert \cite{Ruppert90} gives an effective computational method for determining when an abelian surface $X$  (a rank-2 abelian variety) is isomorphic or isogenous to a product of elliptic curves in terms of a period matrix for $X$.

\paragraph{Split Jacobian varieties.}

Examples \ref{ex:d6} and \ref{ex:d2} show that the distinctions between splitting as p.a.v.s and as complex tori are important even in the case of Jacobian varieties. 
Here we discuss a few results about Jacobians and their splitting, ending with a no-go theorem for certain split abelian surfaces.

Let $\Om^1_C$ denote the sheaf of holomorphic one-forms on a smooth genus $g$ Riemann surface $C$. 
The space of global holomorphic one-forms is then $H^0(\Om_C^1)$ and is isomorphic to $\C^g$. 
Now note that any class $[\g]\in H_1(C,\Z)$ gives rise to a map
\begin{gather}
    \g(\w) = \int_\g \w \in \C,
\end{gather}
where $\w\in H^0(\Om_C^1)$. 
It can be shown that the corresponding adjoint map $H_1(C,\Z) \to H^0(\Om_C^1)^*$ is injective, so $H_1(C,\Z)$ is a lattice of rank $2g$ in $H^0(\Om_C^1)^*$. 
We define the {\it Jacobian} of $X$ as
\begin{gather}
    J(C) = H^0(\Om_C^1)^*/H_1(C,\Z) \cong \C^g/\L,
\end{gather}
where $\L$ is a lattice of rank $2g$. Crucially, $J(C)$ inherits a canonical polarization $\Theta_C$ related to the intersection pairing on $C$. 
The divisor whose corresponding line bundle gives rise to $\Theta_C$ is called the $\Theta$-divisor. 
It can be shown \cite{HM63} that the $\Theta$-divisor is always irreducible and, hence, $(J(C),\Theta_C)$ is irreducible (i.e., does not split) as a p.a.v. 

However, there is no obvious obstruction to $(J(C),\Theta_C)$ splitting as a complex torus, as illustrated in example \ref{ex:d6}. It is therefore interesting to consider the inverse problem: which split abelian surfaces $E\times E'$ are isomorphic to the Jacobian of a smooth genus-2 Riemann surface? The following theorem provides a characterization of this problem.

\begin{theorem}\label{H-N}
    {\bf (Hayashida-Nishi \cite{HN65}).} Let $E$ and $E'$ be two elliptic curves whose rings of endomorphisms are both given by the same principal order $\mathfrak o$ of $\Q(\sqrt{-m})$. 
    Then $E\times E'$ can be the Jacobian variety of a genus-2 Riemann surface for all values of $m$, except for $m\in\{1,3,7,15\}$.
\end{theorem}

\noindent Recall that a \emph{principal} (or, \emph{maximal}) order ${\mathfrak o} \subset \Q(\sqrt{-m})$ is $\mathfrak o = \Z[\sqrt{-m}]$ when $m = 1$ or $2$ (mod 4), and $\mathfrak o = \Z[(1+\sqrt{-m})/2]$ when $m = 3$ (mod 4).
($m$ is taken to be square-free, so $m=0$ (mod 4) does not occur unless $m=0$, in which case $\mathfrak o = \Z$.)  
Any order (not necessarily principal) of $\Q(\sqrt{-m})$ is of the form $  \Z+ c\, {\mathfrak o}$ for some positive integer $c$, where ${\mathfrak o}$ is a principal order for $\Q(\sqrt{-m})$.%
\footnote{$c$, the \emph{conductor} of the order $\Z+c{\mathfrak o}$ in the principal order $\mathfrak o$, depends in a slightly complicated way on $\t$, the modulus of the elliptic curve.
If $\t = (p/q) + (r/s) \sqrt{-m} \in \Q(\sqrt{-m})$ with $p,q,r,s$ integers with $\gcd(p,q)=\gcd(r,s)=1$ and $q$ and $s$ positive, then $c= {\rm lcm} (q,s)$ if $m=1$ or $2$ (mod 4), and $c= {\rm lcm}(q',s')$ if $m=3$ (mod 4), where $q'\doteq q/2$ if $2|q$ and $q$ otherwise, and similarly for $s'$.
}

An immediate implication of theorem \ref{H-N} is that neither $E_i \times E_i$ nor $E_\r\times E_\r$ can be the Jacobian variety of a smooth genus-2 Riemann surface. 
However, they can still be isogenous to one. 
We will show this for $E_\r\times E_\r$. 
(Here we are using the notation $E_\t$ for an elliptic curve of modulus $\t$, and have defined $\r \doteq (1+\sqrt{-3})/2 = e^{i\pi/ 3}$.)
    
We start by observing the following chain of isogenies and isomorphisms:
\begin{gather}
    E_\r \sim E_{2\r} \cong E_{2\r-1} = E_{i\sqrt{3}} \sim E_{i/\sqrt{3}}.
\end{gather}
The first and last relation come from the fact that we may replace $\t$ by $p\t$ for any $p\in \N$ to obtain an isogenous elliptic curve, while the middle isomorphism comes  from the periodicity of $\t$. 
Now, by using example \ref{ex:d6} and computing that $c=i$ corresponds to $z= i/\sqrt{3}$ in \eqref{eq:d6}, we obtain
\begin{gather}
    J(X^{i}[D_6])\cong E_{i\sqrt{3}} \times E_{i/\sqrt{3}}\sim E_\r\times E_\r.
\end{gather}
That is, $E_\r\times E_\r$ is isogenous to the Jacobian of the surface defined by $y^2=x^6+1$.

Let us stress that the endomorphism algebras of $E_\r$, $E_{i\sqrt{3}}$ and $E_{i/\sqrt{3}}$ are all $\Q(\sqrt{-3})$. 
However, only $E_\r$ has endomorphism ring $\mathfrak{o}=\Z[\r]$. 
The other two have endomorphism rings given by the non-principal order
\begin{gather}
    \mathrm{End}(E_{i\sqrt{3}}) = \Z[i\sqrt{3}]= \mathrm{End}(E_{i/\sqrt{3}}) 
\end{gather}
As such, the Hayashida-Nishi condition does not apply to $E_{i\sqrt{3}}\times E_{i/\sqrt{3}}$.


\bibliography{split.bib}

\end{document}